\documentclass[a4paper,11pt]{article}
\usepackage{jheppub_mod}
\usepackage[T1]{fontenc}

\usepackage{hyperref}
\usepackage{graphicx}
\usepackage{amsmath,amssymb,mathtools,slashed}
\usepackage{booktabs,tabulary}
\usepackage{dsfont}
\usepackage{color}
\usepackage{multirow}
\usepackage{subcaption}
\usepackage{braket}

\newcommand{\mpi}{M_\pi}

\newcommand{\beq}{\begin{equation}}
\newcommand{\eeq}{\end{equation}}
\newcommand{\diff}{\text{d}}
\newcommand{\eps}{\epsilon}

\newcommand{\Order}{\mathcal{O}}
\newcommand{\M}{\mathcal{M}}

\newcommand{\Lagr}{\mathcal{L}}

\newcommand{\TeV}{\,\text{TeV}}
\newcommand{\GeV}{\,\text{GeV}}
\newcommand{\MeV}{\,\text{MeV}}

\renewcommand{\Im}{\text{Im}\,}
\newcommand{\Imgg}{\text{Im}_{\gamma\gamma}\,}
\renewcommand{\Re}{\text{Re}\,}

\newcommand{\A}{\mathcal{A}}

\newcommand{\etapp}{\eta^{(\prime)}}
\newcommand{\sm}{s_\text{m}}
\newcommand{\Br}{\text{Br}}
\newcommand{\ml}{m_\ell}

\allowdisplaybreaks[1]

\title{Improved Standard-Model predictions for $\boldsymbol{\eta^{(\prime)}\to \ell^+  \ell^-}$}

\author[a]{Noah Messerli,}
\author[a]{Martin Hoferichter,}
\author[b]{Bai-Long Hoid,}
\author[a]{Simon Holz,}
\author[c]{and \mbox{Bastian Kubis}}

\affiliation[a]{
Albert Einstein Center for Fundamental Physics, Institute for Theoretical Physics, University of Bern, Sidlerstrasse 5, 3012 Bern, Switzerland}

\affiliation[b]{
Institut f\"ur Kernphysik and PRISMA$^+$  Cluster of Excellence, Johannes Gutenberg Universit\"at,  55099 Mainz, Germany} 

\affiliation[c]{
Helmholtz-Institut f\"ur Strahlen- und Kernphysik (Theorie) and \\
Bethe Center for Theoretical Physics, Universit\"at Bonn, 53115 Bonn, Germany}

\emailAdd{noah.messerli@unibe.ch}
\emailAdd{hoferichter@itp.unibe.ch}
\emailAdd{holz@itp.unibe.ch}
\emailAdd{lonbai@uni-mainz.de}
\emailAdd{kubis@hiskp.uni-bonn.de}

\abstract{The rare decays $\eta^{(\prime)}\to\ell^+\ell^-$, $\ell\in\{e,\mu\}$, are highly suppressed in the Standard Model, both by their chirality structure and the required loop attaching the lepton line to the $\eta^{(\prime)}\to\gamma^*\gamma^*$ matrix element. The latter is described by a single scalar function, the transition form factor, which has recently been studied in great detail for $\eta^{(\prime)}$ in the context of the pseudoscalar-pole contributions to hadronic light-by-light scattering in the anomalous magnetic moment of the muon. Based on these results, we evaluate the corresponding prediction for the $\eta^{(\prime)}$ dilepton decays, supplemented by an improved evaluation of the asymptotic contributions including pseudoscalar mass effects. In particular, the dispersive representation for the $\eta^{(\prime)}$ transition form factors allows us, for the first time, to perform a robust evaluation of the imaginary parts due to subleading channels besides the dominant two-photon cut.  
Our final results are $\text{Br}[\eta\to e^+e^-]=5.37(4)(2)[4]\times 10^{-9}$, $\text{Br}[\eta\to \mu^+\mu^-]=4.54(4)(2)[4]\times 10^{-6}$, $\text{Br}[\eta'\to e^+e^-]=1.80(2)(3)[3]\times 10^{-10}$, and $\text{Br}[\eta'\to \mu^+\mu^-]=1.22(2)(2)[3]\times 10^{-7}$, where the errors refer to the uncertainty in the normalized branching fraction, the one propagated from $\text{Br}[\eta^{(\prime)}\to\gamma\gamma]$, and the total uncertainty, respectively.  
The branching fraction for $\eta\to\mu^+\mu^-$ exhibits a mild $1.6\sigma$ tension with experiment, and we explore the bounds that can be derived on physics beyond the Standard Model.}

\begin{document} 

\maketitle

\section{Introduction}
\label{sec:intro}

\begin{figure}[t]
    \centering
    \includegraphics[width=0.6\linewidth]{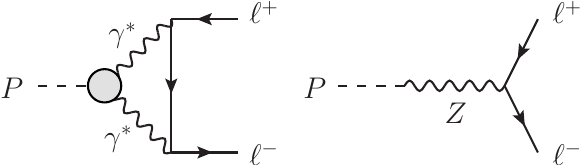}
    \caption{Dilepton decay $P\to\ell^+\ell^-$, $P=\pi^0,\eta,\eta^\prime$, $\ell=e,\mu$, in the SM, via two-photon exchange (left) and $Z$-boson exchange (right). The gray circle denotes the pseudoscalar TFF into two virtual photons. }
    \label{fig:diagrams}
\end{figure}

Dilepton decays of pseudoscalar mesons are promising probes of physics beyond the Standard Model (BSM), since the absence of (pseudo-)scalar currents in the SM entails a suppression of the decay rate with the lepton mass, and the resulting helicity suppression may be lifted in BSM scenarios~\cite{Soni:1974aw}.  Moreover, for light pseudoscalars the direct contribution from $Z$-exchange is small~\cite{Arnellos:1981bk,Masjuan:2015lca,Masjuan:2015cjl}, leaving two-photon exchange as the dominant decay mechanism; see Fig.~\ref{fig:diagrams}. Combining the resulting loop and chirality suppression, the generic size of the branching fraction can be estimated as~\cite{Drell:1959}
\beq
\label{scaling}
\frac{\Br[P\to \ell^+\ell^-]}{\Br[P\to \gamma\gamma]}\simeq
\Big(\frac{\alpha}{\pi}\Big)^2\frac{\ml^2}{M_P^2} \pi^2 \log^2\frac{\ml}{M_P}\simeq \begin{cases}
    2\times 10^{-8} & \qquad\pi^0\to e^+e^-,\\
     2\times 10^{-9} & \qquad\eta\to e^+e^-,\\
     5\times 10^{-6} & \qquad\eta\to \mu^+\mu^-,\\
    9\times 10^{-10} & \qquad\eta^\prime\to e^+e^-,\\
     3\times 10^{-6} & \qquad\eta^\prime\to \mu^+\mu^-.\\
     \end{cases}
\eeq
In practice, this scaling tends to capture the right order of magnitude, but the detailed prediction depends crucially on the matrix element in the two-photon-exchange diagram, i.e., the transition form factor (TFF) describing the amplitude for $P\to\gamma^*\gamma^*$. For instance, in the case of the $\pi^0$, the resulting branching fraction~\cite{Hoferichter:2021lct}
\beq
\label{BR_piee}
\Br[\pi^0\to e^+e^-]=6.25(3)\times 10^{-8}
\eeq
exceeds the naive estimate~\eqref{scaling} by about a factor of three. With a SM branching fraction at the level of $10^{-8}$, plausible BSM scenarios in which the loop (and chirality) suppression is lifted---mediated by pseudoscalar/axial-vector effective operators in SM effective field theory (SMEFT)~\cite{Grzadkowski:2010es,Buchmuller:1985jz} or new light degrees of freedom such as axial-vector $Z'$
bosons~\cite{Kahn:2007ru,Kahn:2016vjr} or
axion-like
particles~\cite{Chang:2008np,Andreas:2010ms,Bauer:2017ris,Alves:2017avw,
Altmannshofer:2019yji,Bauer:2021mvw}---could reveal themselves when confronting precision measurements of $P\to\ell^+\ell^-$ decays with their SM prediction. In this regard, the chirality structure of pseudoscalar contributions can yield sizable enhancement factors, especially for the electron modes. 

For the $\pi^0$, the corresponding SM test goes back as far as Ref.~\cite{Soni:1974aw}, and for a while the KTeV measurement~\cite{Abouzaid:2006kk} indeed suggested a $3\sigma$ tension with the SM prediction~\eqref{BR_piee}. However, when using improved radiative corrections~\cite{Vasko:2011pi,Husek:2014tna}
beyond a point-like $\pi^0\to e^+e^-$ vertex~\cite{Bergstrom:1982wk}, the result,
\beq
\label{KTeV_BR}
\Br[\pi^0\to e^+e^-]\big|_\text{KTeV}=6.85(27)(23)\times 10^{-8},
\eeq
agrees with Eq.~\eqref{BR_piee} at the level of $1.7\sigma$,\footnote{By convention, the comparison between theory and experiment is performed at the level of the leading-order QED result, so that radiative corrections need to be subtracted from the experimental measurement, which typically requires an extrapolation in the photon cut.}
and the resulting BSM constraints are derived in Ref.~\cite{Hoferichter:2021lct}, probing scales up to $4\TeV$ in the case of a pseudoscalar operator. In the meantime, also a preliminary result from NA62 has become available~\cite{Boboc:2024afc}
\beq
\label{NA62_BR}
\Br[\pi^0\to e^+e^-]\big|_\text{NA62}=6.22(39)\times 10^{-8},
\eeq
whose central value almost coincides with Eq.~\eqref{BR_piee}. Moreover, it is striking that the uncertainty of the theoretical prediction exceeds the current experimental sensitivity by an order of magnitude, presenting an opportunity for an even more stringent SM test in the future.

To arrive at a SM prediction at this level of precision, a detailed understanding of the $\pi^0\to\gamma^*\gamma^*$ TFF is required, as had been developed in the context of the pion-pole contribution~\cite{Hoferichter:2018dmo,Hoferichter:2018kwz,Hoferichter:2025fea} in a dispersive approach to hadronic light-by-light (HLbL) scattering~\cite{Hoferichter:2013ama,Colangelo:2014dfa,Colangelo:2014pva,Colangelo:2015ama,Colangelo:2017qdm,Colangelo:2017fiz,Danilkin:2021icn,Ludtke:2023hvz,Hoferichter:2023tgp,Hoferichter:2024fsj,Deineka:2024mzt,Hoferichter:2024vbu,Hoferichter:2024bae} based on the solution of Khuri--Treiman equations~\cite{Khuri:1960zz} for $\gamma^*\to 3\pi$~\cite{Niecknig:2012sj,Schneider:2012ez,Hoferichter:2012pm,Hoferichter:2014vra}, with further applications to hadronic vacuum polarization~\cite{Hoferichter:2019gzf,Hoid:2020xjs,Hoferichter:2023bjm,Hoferichter:2025lcz} and electroweak corrections~\cite{Ludtke:2024ase,Hoferichter:2025yih}; see Refs.~\cite{Aoyama:2020ynm,Aliberti:2025beg,Hertzog:2025ssc} for reviews. While consistent with earlier calculations in chiral perturbation theory (ChPT)~\cite{Savage:1992ac,GomezDumm:1998gw}, vector-meson-dominance (VMD) approaches~\cite{Ametller:1993we,Knecht:1999gb,Silagadze:2006rt,Husek:2015wta}, Canterbury approximants~\cite{Masjuan:2015lca,Masjuan:2017tvw}, and Dyson--Schwinger equations~\cite{Weil:2017knt,Eichmann:2017wil}, as well as first results from lattice QCD~\cite{Christ:2022rho}, a dispersive approach allows one to profit from all available low-energy data both in the space- and time-like domain, to predict the doubly-virtual behavior of the TFF from singly-virtual data, and to implement a smooth matching to short-distance constraints, leading to a precise result with fully controlled uncertainties~\cite{Hoferichter:2018dmo,Hoferichter:2018kwz}. In deriving these dispersive representations, low-energy space-like data for the TFF were not included as a constraint, so that the recent measurements by BESIII~\cite{BESIII:2025zjx} and A2~\cite{Prakhov:2025ejz} yield powerful consistency checks. In addition, in Ref.~\cite{Hoferichter:2021lct} the comparison to earlier approaches using a dispersion relation in the pseudoscalar mass squared~\cite{Bergstrom:1983ay,Ametller:1983ec,Dorokhov:2007bd,Dorokhov:2008cd,
Dorokhov:2009xs} is studied, emphasizing the resulting model dependence due to the choice of the interpolating field and the restriction to the two-photon cut, while observing, empirically, that the numerical result in the case of $\pi^0\to e^+e^-$ remains close to the full calculation due to the small pseudoscalar mass and the dominance of the two-photon intermediate state.

In this work, we generalize the same program to dilepton decays of $\etapp$, profiting from the detailed studies on the corresponding TFFs again in the context of pseudoscalar poles in HLbL scattering~\cite{Stollenwerk:2011zz,Hanhart:2013vba,Kubis:2015sga,Holz:2015tcg,Holz:2022hwz,Holz:2022smu,Holz:2024lom,Holz:2024diw}. In these cases, the experimental situation can be summarized as
\begin{align}
 \Br[\eta\to e^+e^-]&<7.0\times 10^{-7}\quad  \text{at }90\%\text{ CL} && \text{\cite{SND:2018egu}},\notag\\
 \Br[\eta\to \mu^+\mu^-]&=5.8(8)\times 10^{-6}\qquad &&\text{\cite{ParticleDataGroup:2024cfk,Abegg:1994wx,Dzhelyadin:1980kj}},\notag\\
 \Br[\eta^\prime\to e^+e^-]&<5.6\times 10^{-9}\quad  \text{at }90\%\text{ CL} && \text{\cite{Achasov:2015mek}},
\end{align}
while no limits for $\eta^\prime \to\mu^+\mu^-$ are available at present. Starting from a brief overview of the formalism for pseudoscalar dilepton decays in Sec.~\ref{sec:formalism}, we summarize the main features of the dispersive TFFs in Sec.~\ref{sec:low}. One key difference compared to the $\pi^0$ concerns the fact that the doubly-virtual TFF is no longer predicted from singly-virtual input alone due to the different isospin structure, another the higher pseudoscalar mass and thus the increased importance of imaginary parts besides those due to the two-photon cut. Especially for the $\eta^\prime$, intermediate states such as $\pi^+\pi^-\gamma$ are found to play a relevant role, leading to corrections to the naive unitarity bound~\cite{Berman:1960zz,Pratap:1972tb} much larger than observed in the case of $K_L\to\ell^+\ell^-$~\cite{Hoferichter:2023wiy}.
Another consequence of the larger pseudoscalar mass is that mass corrections to the asymptotic contributions~\cite{Hoferichter:2020lap,Zanke:2021wiq} should be considered, novel results for which are presented in Sec.~\ref{sec:asym}. Combined with the small $Z$-boson contribution, our final results for the SM predictions are given in Sec.~\ref{sec:SM}, leading to the BSM constraints in Sec.~\ref{sec:BSM}. We conclude in Sec.~\ref{sec:summary}.

%---------------------------------------------------------------------------------------------------

\section{Formalism}
\label{sec:formalism}

The normalized branching fraction for the two-photon contribution to the decay $P\to\ell^+\ell^-$ can be expressed as
\begin{align}
\frac{\Br[P\to \ell^+\ell^-]}{\Br[P\to\gamma\gamma]}&=2\sigma_\ell(q^2)\Big(\frac{\alpha}{\pi}\Big)^2\frac{\ml^2}{M_P^2}\big|\A_\ell(q^2)\big|^2,\qquad
\sigma_\ell(q^2)=\sqrt{1-\frac{4\ml^2}{q^2}},\notag\\
\A_\ell(q^2)&=\frac{2i}{\pi^2q^2}\int\diff^4 k\frac{q^2k^2-(q\cdot k)^2}{k^2(q-k)^2[(p-k)^2-\ml^2]}\tilde F_{P\gamma^*\gamma^*}\big(k^2,(q-k)^2\big), \label{Eq:redamp_def}
\end{align}
where $\A_\ell$ denotes the reduced amplitude, the variable $q^2=M_P^2$ is fixed at the mass of the pseudoscalar meson, $\alpha=e^2/(4\pi)$, and the TFF enters normalized to its on-shell value
\beq
\tilde F_{P\gamma^*\gamma^*}\big(q_1^2,q_2^2\big)=\frac{F_{P\gamma^*\gamma^*}\big(q_1^2,q_2^2\big)}{F_{P\gamma\gamma}},\qquad F_{P\gamma\gamma}\equiv F_{P\gamma^*\gamma^*}(0,0),
\eeq
which relates to the two-photon decay width via
\beq
\label{ggwidth}
\Gamma[P\to\gamma\gamma]=\frac{\pi\alpha^2M_P^3}{4}F_{P\gamma\gamma}^2.
\eeq
For the TFF, we follow the convention
\beq
i\int \diff^4x \, e^{iq_1\cdot x} \, \langle 0 \vert T \{ j_\mu(x) \, j_\nu(0)\} \vert P(q_1+q_2) \rangle
= \epsilon_{\mu\nu\alpha\beta} \, q_1^\alpha \, q_2^\beta \, F_{P\gamma^*\gamma^*}(q_1^2,q_2^2),
\label{Eq:TFF_matel}
\eeq
with electromagnetic current $j_\mu(x)$ and $\eps^{0123}=+1$. This sign convention matters to ensure consistency with the $Z$-boson contribution~\cite{Masjuan:2015cjl}
\beq
\label{SM_Z}
\A^Z_\ell=-\frac{G_F}{\sqrt{2}\,\alpha^2}\bigg\{\frac{F_\pi}{F_{\pi\gamma\gamma}},
\frac{\sqrt{2}F_\eta^8-F_\eta^0}{\sqrt{6}\,F_{\eta\gamma\gamma}},
\frac{\sqrt{2}F_{\eta^\prime}^8-F_{\eta^\prime}^0}{\sqrt{6}\,F_{\eta^\prime\gamma\gamma}}\bigg\},
\eeq
for $\pi^0$, $\eta$, and $\eta^\prime$, respectively, where the singlet and octet decay constants $F_P^{0,8}$ are defined in the standard conventions~\cite{Gan:2020aco}
 \beq
 \label{F0_F8_def}
\begin{pmatrix}
    F_\eta^8 & F_\eta^0\\
    F_{\eta'}^8 & F_{\eta'}^0
\end{pmatrix}
\equiv\begin{pmatrix}
F_8 \cos \theta_8& -F_0 \sin \theta_0\\
F_8\sin\theta_8 & F_0\cos\theta_0
\end{pmatrix},
 \eeq
and $G_F$ denotes the Fermi constant as determined in muon decay~\cite{MuLan:2012sih}. Numerically, one finds that 
\beq
\label{SM_Z_num}
\A^Z_\ell[\eta]=-0.032,\qquad \A^Z_\ell[\eta^\prime]=0.032,
\eeq
which are insensitive to the input parameters at the level of precision set by the two-photon contribution and thus the $\etapp$ TFFs. 
For their dispersive reconstruction, we follow the conventions from Refs.~\cite{Holz:2024lom,Holz:2024diw}, deferring the discussion of the main features of the resulting dispersive representation to Sec.~\ref{sec:low}.

Thanks to the normalization to the two-photon mode, the corresponding imaginary part
\beq
\Imgg \mathcal{A}_\ell(q^2)=\frac{\pi}{2\sigma_\ell(q^2)}\log \big[y_\ell(q^2)\big],\qquad
y_\ell(q^2)=\frac{1-\sigma_\ell(q^2)}{1+\sigma_\ell(q^2)},
\eeq
does not depend on TFF properties anymore; numerically, one obtains
\beq
\label{Imgg}
\Imgg \mathcal{A}_\ell(q^2)=\begin{cases}
    -17.52& \qquad\pi^0\to e^+e^-,\\
     -21.92& \qquad\eta\to e^+e^-,\\
     -5.47 & \qquad\eta\to \mu^+\mu^-,\\
    -23.68 & \qquad\eta^\prime\to e^+e^-,\\
     -7.06 & \qquad\eta^\prime\to \mu^+\mu^-,\\
     \end{cases}
\eeq
with negligible uncertainties from the particle masses~\cite{ParticleDataGroup:2024cfk}. In the absence of other cuts, this imaginary part yields a unitarity bound~\cite{Berman:1960zz,Pratap:1972tb}
\beq
\big|\A_\ell(q^2)\big|^2\geq \big|\Imgg \mathcal{A}_\ell(q^2)\big|^2,
\eeq
which is exact in the case of the $\pi^0$, while for $\etapp$ corrections arise, primarily due to $2\pi\gamma$ and $3\pi\gamma$ cuts.

To reconstruct the real part of the isovector component of the $\etapp$ TFFs, we use a double-spectral representation of the form
\beq
\tilde F_{P\gamma^*\gamma^*}(q_1^2,q_2^2)=\frac{1}{\pi^2}\int_{4\mpi^2}^{\Lambda^2}\diff x\int_{4\mpi^2}^{\Lambda^2}\diff y\frac{\tilde \rho(x,y)}{(x-q_1^2)(y-q_2^2)}+ (q_1 \leftrightarrow q_2),
\eeq
where $\tilde\rho(x,y)$ denotes the (normalized) double-spectral density and $\Lambda$ an integration cutoff above which an asymptotic contribution needs to be added. Accordingly, one is led to a representation of the reduced amplitude in terms of $\tilde\rho(x,y)$
\beq
\label{Al_disp}
\A_\ell^\text{disp}(q^2)=\frac{2}{\pi^2}
 \int_{4\mpi^2}^{\Lambda^2} \diff x \int_{4\mpi^2}^{\Lambda^2} \diff y\frac{\tilde \rho(x,y)}{x y} K_\ell^\text{disp}(x,y),
\eeq
where the kernel function~\cite{Hoferichter:2021lct}
\begin{align}
 K_\ell^\text{disp}(x,y)&=\frac{2i}{\pi^2 q^2}\int \diff^4 k
 \frac{q^2k^2-(q\cdot k)^2}{k^2(q-k)^2[(p-k)^2-\ml^2]}\frac{x y}{(k^2-x)[(q-k)^2-y]} \label{Eq:kernel_func}\\
&=\frac{1}{2q^2}\Big(x \bar B_0(y,\ml)+y \bar B_0(x,\ml)\Big)+L(x,y)-L(x,0)-L(0,y)+L(0,0),\notag\\
L(x,y)&=\frac{\lambda(x,y,q^2)}{2q^2}C_0(q^2,\ml,x,y),\qquad \lambda(x,y,z)=x^2+y^2+z^2-2(x y+x z+y z),\notag
\end{align}
can be expressed in terms of the standard loop functions
\begin{align}
\label{loop_functions}
 \bar B_0(x,\ml)&=\frac{1}{i\pi^2}\int\frac{\diff^4 k}{(k^2-x)\big((p-k)^2-\ml^2\big)}-(x\to 0)\notag\\
 &=-\int_0^1\diff u\log\bigg[1+\frac{x}{\ml^2}\frac{1-u}{u^2}\bigg]
 =\frac{x}{2\ml^2}\bigg[\log\frac{\ml^2}{x}-\sigma_\ell(x)\log \big[y_\ell(x)\big]\bigg],\notag\\
 C_0(q^2,\ml,x,y)&=\frac{1}{i\pi^2}\int\frac{\diff^4 k}{(k^2-x)\big((q-k)^2-y\big)\big((p-k)^2-\ml^2\big)}\notag\\
 &=-\int_0^1\diff u\int_0^{1-u}\diff v\, \big[\Delta(x,y,u,v)\big]^{-1},\notag\\
 \Delta(x,y,u,v)&=u x+v y-u vq^2+(1-u-v)^2\ml^2.
\end{align}
As a special case, the representation~\eqref{Al_disp} reproduces a VMD approximation $\tilde\rho(x,y)=\pi^2x\delta(x-M_V^2)y\delta(y-M_V^2)$, as required for the isoscalar and effective-pole components of the TFF, see Secs.~\ref{sec:isoscalar} and~\ref{sec:effective}, while the full representation will be used for the dominant isovector contribution discussed in Sec.~\ref{sec:isovector}.

Finally, the asymptotic contribution~\cite{Holz:2024lom,Holz:2024diw}
\beq
\label{asym}
\tilde F_{P\gamma^*\gamma^*}^\text{asym}(q_1^2,q_2^2)
 = \frac{\bar F^P_\text{asym}}{F_{P\gamma\gamma}}\int_{\sm}^\infty \diff x \frac{q_1^2q_2^2}{(x-q_1^2)^2(x-q_2^2)^2},
\eeq
expressed in terms of a transition point $\sm$ and an asymptotic coefficient $\bar F^P_\text{asym}$, e.g., $\bar F^{\pi^0}_\text{asym} = 2F_\pi$,
requires a different integration kernel, leading to the representation~\cite{Hoferichter:2021lct}
\beq
\label{A_ell_asym}
\A^\text{asym}_\ell(q^2)=-\frac{\bar F^P_\text{asym}}{F_{P\gamma\gamma}}\int_{\sm}^\infty \diff x
 \int_0^1\diff u\int_0^{1-u}\diff v\, uv\bigg[\frac{3}{[\Delta(x,x,u,v)]^2}+\frac{(1-u-v)^2\big(q^2-4\ml^2\big)}{[\Delta(x,x,u,v)]^3}\bigg].
\eeq
However, this result only applies in the limit of vanishing mass corrections, which at least for the $\eta^\prime$ does not present a viable approximation anymore~\cite{Holz:2024lom,Holz:2024diw}. In Sec.~\ref{sec:asym} we will therefore construct the generalization of Eq.~\eqref{A_ell_asym} including pseudoscalar mass corrections.

\section{Low-energy contributions}
\label{sec:low}

The $\etapp$ TFFs, defined through the matrix element in Eq.~\eqref{Eq:TFF_matel}, have been studied in the context of the HLbL scattering contribution to the muon $g-2$~\cite{Holz:2024lom,Holz:2024diw,Masjuan:2017tvw,ExtendedTwistedMass:2022ofm,Gerardin:2023naa,Estrada:2024cfy}. Following Refs.~\cite{Holz:2024lom,Holz:2024diw}, the normalized TFFs are decomposed as
\begin{equation}
    \tilde{F}_{\etapp \gamma^*\gamma^*}(q_1^2,q_2^2) = \tilde{F}_{\etapp}^{(I=1)}(q_1^2,q_2^2) + \tilde{F}_{\etapp}^{(I=0)}(q_1^2,q_2^2) + \tilde{F}_{\etapp}^{\text{eff}}(q_1^2,q_2^2) + \tilde{F}_{\etapp}^{\text{asym}}(q_1^2,q_2^2), \label{Eq:tff_compl}
\end{equation}
where the vanishing isospin $I=0$ of $\etapp$ leads to two separate low-energy contributions, both photons in the final state carrying isovector quantum numbers, $\tilde{F}_{\etapp}^{(I=1)}$, or isoscalar ones, $\tilde{F}_{\etapp}^{(I=0)}$. The effective poles, $\tilde{F}_{\etapp}^{\text{eff}}$ serve the interpolation to the asymptotic region as well as imposing the normalizations from experiment~\cite{ParticleDataGroup:2024cfk}
\beq
\label{F_etagg_etapgg}
F^\text{exp}_{\eta\gamma\gamma}=0.2736(48)\GeV^{-1},\qquad
F^\text{exp}_{\eta'\gamma\gamma}=0.3437(55)\GeV^{-1}.
\eeq
In contrast to the $\pi^0$, the comparison to anomaly predictions~\cite{Wess:1971yu,Witten:1983tw} is not immediate due to $\etapp$ mixing, and we use results 
derived from $e^+e^-\to e^+e^-\eta$~\cite{JADE:1985biu,CrystalBall:1988xvy,Roe:1989qy,Baru:1990pc,KLOE-2:2012lws} and the global $\eta^\prime$ fit of the Review of Particle Physics (RPP)~\cite{ParticleDataGroup:2024cfk}.\footnote{For the $\eta^\prime$, the RPP global fit and direct average of data from $e^+e^-\to e^+e^-\eta'$~\cite{CrystalBall:1988xvy,TPCTwoGamma:1988izb,Roe:1989qy,Butler:1990vv,Baru:1990pc,CELLO:1990klc,CrystalBall:1991zkb,L3:1997ocz} differ in their central values, but are consistent within uncertainties.}
The first three contributions in Eq.~\eqref{Eq:tff_compl} are the subject of discussion in this section, while the asymptotic contribution, $\tilde{F}_{\etapp}^{\text{asym}}$, incorporating the leading asymptotic behavior from the light-cone expansion, is discussed in Sec.~\ref{sec:asym}.

\subsection{Isovector transition form factor}
\label{sec:isovector}

In Refs.~\cite{Holz:2024lom,Holz:2024diw}, the dominant low-energy pieces are reconstructed dispersively via their $2\pi$ singularities. Accordingly, as a starting point, the decay amplitudes for the process $\etapp \to 2(\pi^+ \pi^-)$ are constructed based on a hidden-local-symmetry ansatz~\cite{Guo:2011ir}. Factorization-breaking effects in the resulting photon virtualities, $q_1^2$ and $q_2^2$, are implemented by taking into account diagrams involving exchange of the $a_2(1320)$ tensor meson as a left-hand-cut contribution, for which a description by means of phenomenological Lagrangian models~\cite{Bando:1987br,Giacosa:2005bw,Ecker:2007us} was constructed. 

Serving dispersive unitarization, an inhomogeneous Muskhelishvili--Omn\`es problem needed to be solved. Approximating the effects of final-state interaction, pairwise rescattering of the final-state pions was introduced by means of the Omn\`es function~\cite{Omnes:1958hv} based on the pion--pion $P$-wave phase shift.
The solution strategy in this case was inspired by the methods developed in Ref.~\cite{Gasser:2018qtg}, which involves the deformation of the path of integration of the kinematic variables in the dispersion integral into the complex plane, using as input a suitably constructed ChPT-based $\pi\pi$ $P$-wave amplitude that allows for a straightforward analytic continuation~\cite{Gasser:1983yg,Bijnens:1997vq,Dax:2018rvs,Niehus:2020gmf,Niehus:2021iin}. The resulting $\etapp \to \pi^+ \pi^- \gamma^*$ partial-wave amplitude is found as
\begin{equation}
    \mathcal{F}_{\etapp \pi\pi\gamma}(t,k^2) = \frac{1}{96 \pi^2} \int_{4 M_\pi^2}^{\Lambda^2} \diff x\, \frac{x \sigma_\pi^3(x) \big[F_\pi^V(x)\big]^*}{x-k^2-i\eps} 
    \Big[f_1(t,x)\Omega(x) + f_1(x,t) \Omega(t) \Big], \label{Eq:etapipig}
\end{equation}
where $t$ is the $\pi\pi$ invariant mass squared, $k^2$ is the photon virtuality, $\sigma_\pi(s)=\sqrt{1-4 \mpi^2/s}$ the $2\pi$ phase-space factor, $F_\pi^V(s)$ the pion vector form factor, $\Omega(s)$ the Omn\`es function~\cite{Omnes:1958hv}, and $f_1(s,t)$ the partial-wave amplitude of the decay into four pions. Furthermore, $\Lambda^2$ is the cutoff, up to which the dispersion will be carried out and which is varied between $\Lambda^2 \in \{1.5,\, 2.5\}\GeV^2$. The dispersion relation above is kept unsubtracted to ensure the correct asymptotic behavior. However, this implies that the sum rule for the TFF normalization, $F_{\etapp \gamma \gamma}$, is violated at a level around $10\, \%$, which is remedied by introduction of the effective-pole term described in Sec.~\ref{sec:effective}.

As input for the pion vector form factor, the representation $F_\pi^V(s) = P(s) \Omega(s)$ was utilized, where the coefficient of the linear polynomial $P(s)$ was fit to the data of Ref.~\cite{Belle:2008xpe}. Constraining the decay amplitude into four pions, $f_1(s,t)$, to behave as $\mathcal{O}(p^6)$ as required by chiral arguments, the representation in Eq.~\eqref{Eq:etapipig} then contains only two free parameters, one associated with the momentum dependence and normalization and another one describing the relative strength of the left-hand-cut contribution, by means of coupling constants in the $a_2(1320)$-exchange diagrams. Both of these parameters are fixed by fits to data of the real photon decay spectra $\etapp \to \pi^+ \pi^- \gamma$, provided by KLOE~\cite{KLOE:2012rfx} for the case of the $\eta$ and BESIII~\cite{BESIII:2017kyd} for the case of the $\eta'$.

\begin{figure}[t]
    \centering
    \begin{minipage}{0.45\linewidth}
        \includegraphics[width=\linewidth]{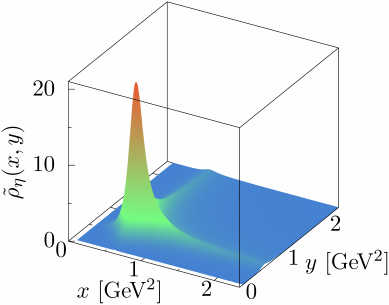}
    \end{minipage}
    \begin{minipage}{0.45\linewidth}
        \includegraphics[width=\linewidth]{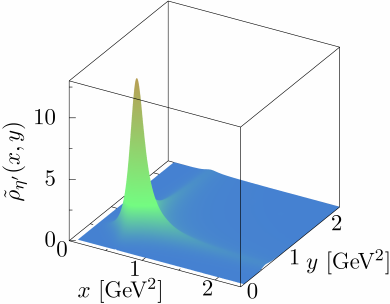}
    \end{minipage}
    \begin{minipage}{0.08\linewidth}
        \centering
        \includegraphics[width=.9\linewidth]{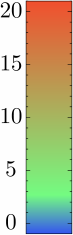}
    \end{minipage}
    \caption{Normalized isovector double-spectral densities of Eq.~\eqref{Eq:doubspecdens} for $\eta$ (left) and $\eta'$ (right).}
    \label{fig:specfunc}
\end{figure}

Utilizing the dominant two-pion intermediate states, another dispersion relation in the Mandelstam variable $t$ can be applied on top of the representation in Eq.~\eqref{Eq:etapipig} in order to determine the isovector part of the TFFs. However, in order to preserve the correct asymptotic behavior, and in the HLbL application of Refs.~\cite{Holz:2024lom,Holz:2024diw} to facilitate the analytic continuation to the  space-like region, as well as for the application in dilepton decays in this work, it is advantageous to express this dispersion relation in a double spectral representation,
\begin{equation}
    \tilde{F}_{\etapp}^{(I=1)}(q_1^2,q_2^2) = \frac{1}{\pi^2} \int_{4M_\pi^2}^{\Lambda^2}\diff x\, \int_{4M_\pi^2}^{\Lambda^2} \diff y\,  \frac{ \tilde{\rho}_{\etapp}(x,y)}{(x-q_1^2)(y-q_2^2)} + (q_1 \leftrightarrow q_2), \label{Eq:TFF_I1}
\end{equation}
which has been symmetrized in the photon virtualities and where the normalized double-spectral density is given by
\begin{equation}
    \tilde{\rho}_{\etapp}(x,y) = \frac{x \sigma_\pi^3(x)}{192 \pi F_{\etapp \gamma \gamma}} \Im \Big\{\big[F_\pi^V(x)\big]^*  \mathcal{F}_{\etapp \pi\pi\gamma}(x,y)\Big\}. \label{Eq:doubspecdens}
\end{equation}
This double-spectral density is shown for both $\eta$ and $\eta'$ in Fig.~\ref{fig:specfunc}, where it can be seen that these densities are enhanced in the $\rho$-resonance region, as expected. The corresponding contribution to the reduced amplitude of the dilepton decays, defined in Eq.~\eqref{Eq:redamp_def}, can then be written as
\begin{equation}
    \mathcal{A}^{\text{disp}}_\ell(q^2) =  \frac{2}{\pi^2}\int_{4M_\pi^2}^{\Lambda^2} \diff x\int_{4\mpi^2}^{\Lambda^2} \diff y\, \frac{\tilde{\rho}_{\eta^{(\prime)}}(x,y)}{xy} K_\ell^\text{disp}(x,y),
    \label{Eq:A_disp}
\end{equation}
with the kernel function given in Eq.~\eqref{Eq:kernel_func}, making use of its symmetry property under exchange of its arguments.

\subsection{Isoscalar transition form factor}
\label{sec:isoscalar}

In contrast to the intricate solution of a dispersion relation required to capture the details of the dominant isovector contribution, a simpler description based on narrow resonances suffices for the isoscalar component, exploiting the small widths of the corresponding isoscalar resonances $\omega$ and $\phi$. Both of them only provide a small contribution to the total TFF compared to the isovector component and it is therefore possible to opt for a description by means of a VMD ansatz~\cite{Holz:2024lom,Holz:2024diw}:
\begin{align}
    \tilde{F}_{\eta^{(\prime)}\gamma^*\gamma^*}^{(I=0)}(q_1^2,q_2^2) = \sum\limits_{V\in\{\omega,\, \phi\}} \frac{w_{\eta^{(\prime)}V\gamma} M_V^4}{(M_V^2-q_1^2)(M_V^2-q_2^2)}, \label{Eq:TFF_I0}
\end{align}
where $M_V$ is the mass of the respective vector meson, and $w_{\eta^{(\prime)}V\gamma}$ are the corresponding weight factors, which can be determined phenomenologically from decay widths for the processes $\phi\rightarrow \eta^{(\prime)} \gamma$, $\omega\rightarrow \eta\gamma$, $\eta^\prime\rightarrow \omega\gamma$, and $\phi,  \omega\rightarrow e^+e^-$, with their respective signs fixed by $U(3)$-symmetry considerations~\cite{Hanhart:2013vba,Gan:2020aco}. Employing the decay widths listed in the RPP, the following numerical values are used in this work
\begin{align}
\label{Eq:isosc_weights_values}
w_{\eta\omega\gamma}&=0.099(7), & w_{\eta\phi\gamma}&=-0.188(5),\notag\\
w_{\eta'\omega\gamma}&=0.071(2),& w_{\eta'\phi\gamma}&=0.155(4).
\end{align}
Employing the zero-width VMD representation of the isoscalar TFF in Eq.~\eqref{Eq:TFF_I0}, the corresponding contribution to the dilepton reduced amplitude can be written as
\begin{equation}
    \mathcal{A}_\ell^{(I=0)}(q^2) = \sum\limits_{V\in\{\omega,\, \phi\}} \mathcal{A}_\ell^V(q^2),\quad \mathcal{A}_\ell^V(q^2) = w_{\etapp V \gamma} K_\ell^\text{disp}(M_V^2,M_V^2).
    \label{Eq:AI0_zerow}
\end{equation}
Going beyond a zero-width approximation, it is possible to describe effects due to the finite but small width by introducing dispersively improved Breit--Wigner factors~\cite{Lomon:2012pn,Moussallam:2013una,Crivellin:2022gfu}. For the case at hand, in Eq.~\eqref{Eq:TFF_I0}, one can accordingly replace~\cite{Zanke:2021wiq}
\begin{equation}
    \frac{M_V^2}{M_V^2-q_i^2} \to \frac{P_V^\text{disp}(q_i^2)}{P_V^\text{disp}(0)}\,
\end{equation}
with
\begin{align}
     P_V^\text{disp}(k^2) &= \frac{1}{\pi} \int_{s_\text{thr}}^{\infty} \diff x\, \frac{\Im P_V^\text{disp}(x)}{x - k^2 - i \eps},\notag\\
     \Im P_V^\text{disp} (k^2) &= \Im \frac{1}{M_V^2-k^2-i \sqrt{k^2} \Gamma_V(k^2)}.
\end{align}
The energy-dependent width of the $\phi$ resonance, as an example, can be parameterized as~\cite{Stamen:2022uqh}
\begin{align}
    \Gamma_\phi(k^2) = \sum\limits_{X_1X_2\in\mathcal{D}} \frac{\gamma_{\phi \to X_1 X_2}(k^2)}{\gamma_{\phi \to X_1 X_2}(M_\phi^2)} \Gamma_{\phi\to X_1 X_2}(k^2) \theta\big( k^2 - (M_{X_1}+M_{X_2})^2 \big)
\end{align}
by means of its decay channels into $\mathcal{D}=\{K^+K^-,\, K^0_S K^0_L,\, \rho \pi\}$  with
\begin{equation}
    \gamma_{\phi\to \rho \pi}(k^2) = \frac{\lambda^{3/2}(k^2,M_\rho^2,M_\pi^2)}{(k^2)^{3/2}},\quad \gamma_{\phi\to \bar{K}K}(k^2) = \frac{(k^2 - 4M_K^2)^{3/2}}{k^2}.
\end{equation}
Furthermore, we introduce centrifugal barrier factors~\cite{VonHippel:1972fg,COMPASS:2015gxz},
\begin{align}
    \Gamma_{\phi \to \bar{K}K}(k^2) &=\Gamma_{\phi \to \bar{K}K} \frac{\sqrt{k^2}}{M_\phi} \frac{M_\phi^2 - 4M_K^2+4p_R^2}{k^2 - 4M_K^2 + 4p_R^2},\quad p_R=202.4\MeV,\\
    \Gamma_{\phi \to \rho \pi}(k^2) &= \Gamma_{\phi \to 3\pi + \rho \pi}, \notag
\end{align}
in order to mitigate the asymptotic behavior of the energy-dependent widths. The constant partial decay widths are fixed to the central values taken from the RPP: $\Gamma_{\phi \to K^+ K^-} = 2.12\MeV$, $\Gamma_{\phi\to K^0_S K^0_L} = 1.43\MeV$, and $\Gamma_{\phi \to 3\pi + \rho \pi} = 0.63\MeV$. Taking into account these finite-width effects, the contribution to the reduced amplitude can be written as
\begin{equation}
    \mathcal{A}_\ell^\phi(q^2) = \frac{w_{\etapp \phi \gamma}}{(\pi P_\phi^\text{disp}(0))^2} \int_{(M_\rho+M_\pi)^2}^\infty \diff x\int_{(M_\rho+M_\pi)^2}^\infty \diff y\, \frac{K^\text{disp}_\ell(x,y)}{xy}\, \Im \big[ P_\phi^\text{disp}(x) \big] \, \Im \big[ P_\phi^\text{disp}(y) \big].
\end{equation}
However, at the targeted level of precision for numerical values of the isoscalar reduced amplitudes, we cannot distinguish results from the zero-width representation and the finite-width one owing to the narrow width of the $\phi$. For the $\omega$ resonance, one can expect a similar behavior. In conclusion, it is safe to say that the isoscalar $\phi$ and $\omega$ resonances can be treated as infinitely narrow for the application to $\etapp$ dilepton decays.

\subsection{Effective poles}
\label{sec:effective}

The effective pole term in the TFF representation of Eq.~\eqref{Eq:tff_compl}, $\tilde{F}_{\etapp}^{\text{eff}}(q_1^2,q_2^2)$, is introduced since the sum of the isovector and isoscalar parts does not saturate the TFF normalization. Additionally, the sum of these two terms does not capture the effects of higher hadronic resonances at virtualities $\gtrsim 1\GeV^2$. The parameters in the effective pole term are then determined by the requirement of the resulting TFFs fulfilling the normalization sum rule exactly, as well as fits to high-energy, space-like data for the processes $e^+ e^- \to e^+ e^- \etapp$~\cite{CELLO:1990klc,CLEO:1997fho,L3:1997ocz,BaBar:2011nrp}. In Refs.~\cite{Holz:2024lom,Holz:2024diw}, two different parameterizations are employed,
 \begin{align}
     \tilde{F}_{\etapp}^{\text{eff}\,(A)}(q_1^2,q_2^2) &= \frac{g_\text{eff}    M_{\text{eff}}^4}{(M_{\text{eff}}^2-q_1^2)(M_{\text{eff}}^2-q_2^2)},\notag\\
     \tilde{F}_{\etapp}^{\text{eff}\,(B)}(q_1^2,q_2^2) &= \sum\limits_{V\in \lbrace \rho',\,\rho''\rbrace} \frac{g_V  M_V^4}{(M_V^2-q_1^2)(M_V^2-q_2^2)},
     \label{Eq:TFF_eff}
 \end{align}
 where in the single-pole variant $(A)$, the effective coupling $g_\text{eff}$ is tuned to restore the normalization. In practice, for both $\eta$ and $\eta'$, its magnitude $|g_\text{eff}|$ is found in a range up to $10\, \%$. Furthermore, in fits to singly-virtual TFF data for $Q^2\geq 5 \GeV^2$, the effective mass parameter is found in the range $(1.3$--$2.2)\, \text{GeV}$. In contrast, in the two-pole variant $(B)$, the mass parameters of the $\rho(1450)\equiv\rho'$ and $\rho(1700)\equiv\rho''$ are fixed to their values listed in the RPP. Then, the coupling of one of these resonances $g_V$ is used to fulfill the normalization sum rule, while the corresponding other coupling is used as a fit parameter.\par
 The resulting contribution to the reduced amplitudes can then, in analogy to Eq.~\eqref{Eq:AI0_zerow}, be expressed as
 \begin{align}
     \mathcal{A}_\ell^{\text{eff}\,(A)}(q^2) &= g_\text{eff} K^\text{disp}_\ell(M_\text{eff}^2,M_\text{eff}^2),\notag\\
     \mathcal{A}_\ell^{\text{eff}\,(B)}(q^2) &= \sum\limits_{V\in \lbrace \rho',\,\rho''\rbrace} g_V K^\text{disp}_\ell(M_V^2,M_V^2).
 \end{align}
 While not being a numerically large effect, the spread between variants $(A)$ and $(B)$ is included in the final uncertainty estimate.

\section{Asymptotic contributions}
\label{sec:asym}

\subsection{Asymptotic form of the transition form factor}

The asymptotic contribution to the $\etapp$ TFFs is described in detail in Refs.~\cite{Holz:2024lom,Holz:2024diw}; here, we first summarize the main results and then discuss the consequences for the loop integral in $\etapp\to\ell^+\ell^-$. Starting point is the leading expansion around the light cone $x^2=0$~\cite{Lepage:1979zb,Lepage:1980fj,Brodsky:1981rp}
\beq
\label{F_asym_BL}
F_{P\gamma^*\gamma^*}(q_1^2,q_2^2)=-\frac{\bar F^P_\text{asym}}{3}\int_0^1\text{d}u\frac{\phi_P(u)}{u q_1^2+(1-u)q_2^2},
\eeq
expressed in terms of a wave function $\phi_P(u)$ that can be expanded into Gegenbauer polynomials.
In the conformal limit~\cite{Braun:2003rp} one has $\phi_P(u)=6u(1-u)$, and also the coefficient $\bar F^P_\text{asym}$ is related to known matrix elements. In the case of the pion, the latter relation reads $\bar F^\pi_\text{asym}=2F_\pi$, with pion decay constant $F_\pi=92.32(10)\MeV$~\cite{ParticleDataGroup:2024cfk}, while for $\etapp$ the coefficients depend on the mixing pattern. We use 
\begin{align}
\label{barF_asym_num}
    \bar F_\text{asym}^\eta&=0.186(7)_\text{norm}(7)_\text{disp}(9)_\text{BL} [13]_\text{tot}\GeV,\notag \\
    \bar F_\text{asym}^{\eta'}&=0.264(5)_\text{norm}(5)_\text{disp}(11)_\text{BL} [13]_\text{tot}\GeV, 
\end{align}
as derived via a superconvergence sum rule in Refs.~\cite{Holz:2024lom,Holz:2024diw}. The uncertainties refer to normalization, systematics of the dispersive representation, and the Brodsky--Lepage (BL) matching, the latter propagated from the singly-virtual space-like data for $e^+e^-\to e^+e^-\etapp$ for $Q^2\geq 5 \GeV^2$. In turn, $\bar F_\text{asym}^{\etapp}$ can be used to determine $\etapp$ mixing angles and decay constants, and in the remainder of this work we will use the corresponding results from Refs.~\cite{Holz:2024lom,Holz:2024diw}, to which we also refer for a comparison to previous determinations~\cite{Escribano:2015yup,Bali:2021qem}. 

The general formula~\eqref{F_asym_BL} implies the limits~\cite{Nesterenko:1982dn,Novikov:1983jt}   
\begin{align}
     \lim\limits_{Q^2\to\infty} Q^2 F_{P \gamma^* \gamma^*}(-Q^2,-Q^2) &= \frac{1}{3}\bar{F}^P_{\text{asym}},\notag\\
\lim\limits_{Q^2\to\infty} Q^2 F_{P \gamma^* \gamma^*}(-Q^2,0) &= \bar{F}^{P}_{\text{asym}},
\end{align}
where the second one goes beyond a strict operator product expansion (OPE)~\cite{Gorsky:1987idk,Manohar:1990hu}, as reflected by a stronger sensitivity to the wave function. For that reason, the dispersive representation for the TFF is constructed in such a way that the doubly-virtual limit is saturated by the asymptotic form, while the singly-virtual behavior is determined by space-like data at large virtuality $Q^2\geq 5\GeV^2$. This is why the asymptotic coefficient follows from the dispersive representation via a superconvergence sum rule. 

In practice, Eq.~\eqref{F_asym_BL} is implemented in the dispersive approach by rewriting  the asymptotic contribution as a dispersion relation~\cite{Khodjamirian:1997tk} and then imposing 
a lower matching scale $\sm$. In particular, a prudent choice of the boundary terms in the evaluation  
of the double-spectral density 
\beq
\label{double_spectral_density}
\rho^\text{asym}(q_1^2,q_2^2)=-\pi^2 \bar F^P_\text{asym}q_1^2q_2^2\delta''(q_1^2-q_2^2)
\eeq
allows one to ensure that indeed the result
\beq
\label{Fasym_massless}
F^\text{asym}_{P\gamma^*\gamma^*}(q_1^2,q_2^2)=\bar F_\text{asym}^P\int_{\sm}^\infty \text{d}x\frac{q_1^2q_2^2}{(x-q_1^2)^2(x-q_2^2)^2}
\eeq
vanishes in the singly-virtual limit. 
The motivation for this procedure is that in this limit the dispersive representation has the same asymptotic behavior as Eq.~\eqref{F_asym_BL}, with a coefficient that can be determined by a fit to space-like TFF data measured in $e^+e^-\to e^+e^- P$. With $\bar F_\text{asym}^P$ thus inferred from the data, via a superconvergence relation, also the doubly-virtual contribution is predicted. 

Once the mass of the pseudoscalar meson becomes comparable to the matching point $\sm$---chosen as $\sm=1.5(3)\GeV^2$ for the $\eta'$ and $\sm=1.4(4)\GeV^2$ for the $\eta$ following the arguments from Refs.~\cite{Holz:2024lom,Holz:2024diw}, e.g., the comparison to light-cone sum rules~\cite{Khodjamirian:1997tk,Agaev:2014wna}---mass corrections should be included, changing  
 Eq.~\eqref{F_asym_BL} to~\cite{Hoferichter:2020lap}
\beq
\label{F_asym_BL_mass}
F_{P\gamma^*\gamma^*}(q_1^2,q_2^2)=-\frac{\bar F^P_\text{asym}}{3}\int_0^1\text{d}u\frac{\phi_P(u)}{u q_1^2+(1-u)q_2^2-u(1-u)M_P^2}.
\eeq
While these corrections do not constitute a complete description of higher-order terms in the light-cone expansion, it appears well motivated to keep the form~\eqref{F_asym_BL_mass}, as it appears naturally in the derivation when keeping the full kinematic relation. 
The corresponding generalization 
 of the double-spectral density~\eqref{double_spectral_density} was derived in Ref.~\cite{Zanke:2021wiq}, and in Ref.~\cite{Hoferichter:2024bae} suitable subtractions were introduced to preserve the behavior $F_{P}^{\text{asym}}(q_1^2,q_2^2)=\Order(q_1^2q_2^2)$ for small virtualities. The resulting representation reads 
 \begin{align}
 \label{Eq:TFF_asym}
     F_{P}^{\text{asym}}(q_1^2,q_2^2) &= \frac{- \bar{F}_{\text{asym}}^{P}}{M_{P}^4} \int_{2 \sm}^{\infty}\text{d}v \bigg[ \frac{q_2^2}{v-q_1^2} \bigg[\frac{1}{v-q_1^2-q_2^2} - \frac{1}{q_1^2-q_2^2} \bigg] f^{\text{asym}}_{P}(v,q_1^2) + (q_1^2 \leftrightarrow q_2^2) \bigg], \notag \\
     f^{\text{asym}}_{P}(v,q^2) &= \frac{(v-2q^2)^2 - M_{P}^2 v}{\sqrt{(v-2q^2)^2 - 2 M_{P}^2 v +M_{P}^4}} +2 q^2 - v,
 \end{align}
and the main challenge in the current application concerns its evaluation inside the $\etapp\to\ell^+\ell^-$ loop integral, to which we turn next. 

\subsection{Implementation in the loop integral}

As a first step towards the implementation including pseudoscalar mass effects, we repeat the calculation of Eq.~\eqref{A_ell_asym} in terms of standard loop functions. Performing the Passarino--Veltman reduction in FeynCalc~\cite{Mertig:1990an,Shtabovenko:2016sxi,Shtabovenko:2020gxv,Shtabovenko:2023idz} and simplifying the result by integration-by-parts identities using FIRE~\cite{Smirnov:2023yhb} (connected via FeynHelpers~\cite{Shtabovenko:2016whf}), we obtain
\begin{align}
\label{F_P_asym_standard_loop}
        \mathcal{A}^{\text{asym}}_\ell &= -\frac{\bar F_\text{asym}^P}{2M_P^2F_{P\gamma\gamma}}\int_{\sm}^\infty \diff x\bigg\{2C_0(M_P^2,\ml,x,x)
        +\frac{1}{\ml^2}\log\frac{x}{\ml^2}\notag\\
        &-
        \frac{x^4 -2x^2\ml^2(5x-M_P^2)+2\ml^4\big[16x^2-6xM_P^2+M_P^4\big] -8 \ml^6 (4x-M_P^2)}{(x - 4 \ml^2) \big[x^2-\ml^2(4x-M_P^2)\big]^2}\notag\\
        &\qquad\times2\tilde B_0\big(\ml^2, \ml^2, x\big)\notag\\
        &+\frac{4 x^3(x-M_P^2) - 
        \ml^2 (4x-M_P^2) \big[4x^2-6xM_P^2+M_P^4\big]}{(4x-M_P^2) \big[x^2-\ml^2(4x-M_P^2)\big]^2}\tilde B_0\big(M_P^2, x, x\big)\notag\\
       &+\frac{8x^2-6xM_P^2+M_P^4}{(4x-M_P^2) \big[x^2-\ml^2(4x-M_P^2)\big]}
        \bigg\},
\end{align}
with loop functions in the conventions of 
Package X~\cite{Patel:2015tea,Patel:2016fam}, i.e., $C_0(M_P^2,\ml,x,x)$ as defined in Eq.~\eqref{loop_functions} coincides with $\texttt{ScalarC0}[\ml^2, \ml^2, M_P^2, \sqrt{x}, \ml, \sqrt{x}]$ and 
\begin{align}
 \tilde B_0\big(\ml^2, \ml^2, x\big)&\equiv   \texttt{DiscB}[\ml^2,\ml,\sqrt{x}]=-\frac{x}{2\ml^2}\sigma_\ell(x)\log \big[y_\ell(x)\big],\notag\\
 \tilde B_0\big(M_P^2, x, x\big)&\equiv   \texttt{DiscB}[M_P^2,\sqrt{x},\sqrt{x}]=2\sqrt{\frac{4x}{M_P^2}-1}\bigg(\arctan\sqrt{\frac{4x}{M_P^2}-1}-\frac{\pi}{2}\bigg);
\end{align}
see Ref.~\cite{Messerli:2025} for more details. 
To generalize this result to the massive case, we need to recast Eq.~\eqref{Eq:TFF_asym} into propagator form, which can be achieved by writing  
\beq
f_P^\text{asym}(v,q^2)=2q^2-v+\frac{1}{2\pi}\int_{-1}^1 \frac{\diff x}{\sqrt{1-x^2}} \frac{(v-2q^2)^2-M^2_Pv}{v_x^+ -q^2},\qquad v_x^\pm=\frac{v\pm M_Px\sqrt{2v-M^2_P}}{2}.
\eeq
In this form, the final result becomes~\cite{Messerli:2025} 
\begin{align}
\label{A_asym_massive}
 \mathcal{A}^{\text{asym}}_{\ell,P} &= \frac{\bar F_{\text{asym}}^P}{64 \pi M_P^4 F_{P \gamma \gamma} }\int_{2\sm}^\infty \diff v \int_{-1}^1 \frac{\diff x}{\sqrt{1-x^2}} \notag\\
 &\times\Bigg\{ b(x,v)B_0\big(\ml^2, \ml^2, v_x^+\big)
 + c_1(x,v) C_0\Big(\ml^2, \frac{M_P^2}{4}, \ml^2 - \frac{M_P^2}{4}, \ml^2, v_x^+, \frac{1}{4} \left(2 v - M_P^2\right)\Big)\notag\\
    &+ c_2(x,v) C_0\big(\ml^2, \ml^2, M_P^2, v_x^+, \ml^2, v_x^+\big) \notag\\
    &+ c(v)\bigg[ 2 C_0\big(\ml^2, \ml^2, M_P^2, 0, \ml^2, v\big) - C_0\Big(\ml^2, \frac{M_P^2}{4}, \ml^2 - \frac{M_P^2}{4}, \ml^2, 0, \frac{1}{4}\left(2v-M_P^2\right)\Big) \notag\\
    &\qquad- C_0\Big(\ml^2, \frac{M_P^2}{4}, \ml^2 - \frac{M_P^2}{4}, \ml^2, v, \frac{1}{4}\left(2v-M_P^2\right)\Big)  \bigg]
\Bigg\},
\end{align}
where the coefficients are defined as
\begin{align}
\label{coefficients}
        b(x,v) &= -\frac{32 x M_P}{v_x^+v_x^-}\sqrt{2v-M_P^2}\Big[v(1-2x^2)+x^2M_P^2\Big],\notag\\
        c_1(x,v) &= \frac{16M_P^2}{v_x^+v_x^-}(x^2-1)(2v-M_P^2)\Big[v(1-2x^2)+x^2M_P^2\Big],\notag\\
        c_2(x,v) &= \frac{16M_P^2}{v_x^+v_x^-}\sqrt{2v-M_P^2}\Big(\sqrt{2v-M_P^2}+2M_P x\Big)\Big[v(1-2x^2)+x^2M_P^2\Big],\notag\\
        c(v) &= \frac{
        32 v^3-308 M_P^2 v^2 + 32 M_P^4 v + M_P^6 }{v^2}.
\end{align}
Since $b(x,v)$ is odd in $x$, all $x$-independent terms in the $B_0$ loop function cancel, which can therefore be reduced to 
\beq
B_0\big(\ml^2, \ml^2, v_x^+\big)\to \texttt{DiscB}[\ml^2,\ml,\sqrt{v_x^+}]-\bigg(1-\frac{v_x^+}{2\ml^2}\bigg)\log\frac{\ml^2}{v_x^+}+\log\frac{v}{v_x^+}. 
\eeq
The $C_0$ functions are given as $C_0(p_1^2,p_2^2,p_3^2,m_1^2,m_2^2,m_3^2)=\texttt{ScalarC0}[p_1^2,p_2^2,p_3^2,m_1,m_2,m_3]$ in Package-X conventions. Finally, we transform the $x$ integration onto $x=\sin t$, to avoid the square-root singularities at $x=\pm 1$.

As a cross check on Eq.~\eqref{A_asym_massive} we considered the limiting case $\ml\to 0$, in which an expansion in $M_P^2$ can be derived, and given subtleties in the numerical evaluation in the case of small lepton masses, this serves as powerful validation of the numerical procedure. That is, singularities for $\ml\to 0$ are present in the individual terms, but they disappear in the sum since the contribution to $c(v)$ cancels altogether for $\ml\to 0$, while $B_0(\ml^2,\ml^2,v_x^+)\to -\log\frac{v_x^+}{v}$.
The expansion of the remainder in $M_P^2$ then has to arrange in such a way that the $1/M_P^4$ in Eq.~\eqref{A_asym_massive} is removed, and indeed we obtain~\cite{Messerli:2025}
\begin{align}
\label{expansion}
    \mathcal{A}^{\text{asym}}_{\ell,P}\Big|_{\ml=0} 
    &= - \frac{\bar F_{\text{asym}}^P }{F_{P\gamma\gamma}}\int_{2\sm}^\infty \frac{\diff v}{v^2}  \bigg[ \frac{1}{2}+\frac{98M_P^2}{90v} + \frac{23M_P^4}{12v^2} + \frac{1427M_P^6}{450v^3} + \frac{4363M_P^8}{840v^4}  + \mathcal{O}\bigg(\frac{M_P^{10}}{v^5}\bigg) \bigg]\notag\\
    &= -\frac{\bar F_{\text{asym}}^P }{F_{P\gamma\gamma}}\frac{1}{\sm}\bigg[\frac{1}{4} + \frac{49M_P^2}{360 \sm}  + \frac{23M_P^4}{288\sm^2} + \frac{1427M_P^6}{28800\sm^3} + \frac{4363M_P^8}{134400\sm^4}+ \mathcal{O}\bigg(\frac{M_P^{10}}{\sm^5}\bigg)\bigg].
\end{align}
In particular, the first term reproduces Eq.~\eqref{A_ell_asym} evaluated in the massless limit $\ml=M_P=0$. The different variants are compared for a representative set of parameters in Table~\ref{tab:asymptotic}. As expected, the mass corrections are most relevant for the $\eta'$, while the effect of finite $\ml$ is small even for the muon channel. Moreover, the result at $\ml=0$ is reproduced accurately from the expansion~\eqref{expansion}, modified to include a finite integration cutoff $\Lambda^2$ in the $v$ integration, up to a small remaining correction in the case of the $\eta^\prime$. The fact that all these limits are correctly reproduced therefore constitutes a valuable consistency check of the general result given in Eq.~\eqref{A_asym_massive}. 

\begin{table}[t]
	\renewcommand{\arraystretch}{1.3}
	    \centering
        	\begin{tabular}{l r r r r}
				   \toprule
    Decay mode & $\mathcal{A}^{\text{asym}}_\ell(q^2)$ & $\mathcal{A}^{\text{asym}}_{\ell,P}(q^2)$ & $\mathcal{A}^{\text{asym}}_{\ell,P}(q^2)\big|_{\ml=0}$ & $\mathcal{A}^{\text{asym}}_{\ell,P}(q^2)\big|_{\ml=0}^{\Order(M_P^8)}$ \\
    \midrule
    $\pi^0 \rightarrow e^+e^-$ & $ -0.0990$ &  $ -0.0993$ & $ -0.0993$ & $ -0.0993$ \\ 
    $\eta \rightarrow e^+e^-$ & $ -0.1278$  &  $-0.1374$ & $ -0.1374$ & $ -0.1374$  \\
    $\eta \rightarrow \mu^+\mu^-$ & $ -0.1264$ &  $ -0.1358$ & $ -0.1374$ & $ -0.1374$\\
    $\eta^\prime \rightarrow e^+e^-$ & $ -0.1487$ &  $ -0.1957$  & $ -0.1957$ & $ -0.1938$ \\
    $\eta^\prime \rightarrow \mu^+\mu^-$ & $ -0.1466$ &  $ -0.1913$ & $ -0.1957$ & $ -0.1938$ \\
    \bottomrule
        \end{tabular}
        \caption{Asymptotic contribution to the reduced amplitude for pseudoscalar dilepton decays for a representative set of parameters: $\sm^{\pi^0} = 1.7\GeV^2$, $\sm^\eta = 1.4\GeV^2$, $\sm^{\eta^\prime} = 1.5\GeV^2$, upper integral cutoff $\Lambda^2 = 10^3\GeV^2$ in the $v$ integration, masses and normalizations from the RPP, and $\bar F_\text{asym}^P$ from Eq.~\eqref{barF_asym_num}. The first column refers to the result without pseudoscalar mass corrections, Eqs.~\eqref{A_ell_asym} or~\eqref{F_P_asym_standard_loop}, the second column to the massive case~\eqref{A_asym_massive}, the third column to its limit for $m_\ell=0$, and the fourth column to the same limit evaluated based on the expansion~\eqref{expansion}.}
        \label{tab:asymptotic}
    \renewcommand{\arraystretch}{1.0}
    \end{table}

\section{Standard-Model predictions}
\label{sec:SM}

\subsection{Final results and uncertainty estimates}

\begin{table}[t]
\renewcommand{\arraystretch}{1.3}
  \centering
\begin{tabular}{lrr}
    \toprule
    Decay mode & $\Re\A_\ell$ & $\Im\A_\ell$ \\
    \midrule 
    $\eta \rightarrow e^+e^-$ & $ 31.25(13)_{\text{disp}}(4)_{\text{BL}}(7)_{\text{asym}}(1)_{\omega,\phi}[15]$ & $-21.836 (3)_{\text{disp}}(0)_{\text{BL}}(0)_{\omega,\phi}[3]$   \\
    $\eta \rightarrow \mu^+\mu^-$ &  $-1.21(7)_{\text{disp}}(3)_{\text{BL}}(7)_{\text{asym}}(0)_{\omega,\phi}[10]$ & $- 5.449 (1)_{\text{disp}}(0)_{\text{BL}}(0)_{\omega,\phi}[1] $  \\
    $\eta' \rightarrow e^+e^-$ & $46.76(12)_{\text{disp}}(18)_{\text{BL}}(10)_{\text{asym}}(2)_{\omega,\phi}[24]$ & $-18.865 (72)_{\text{disp}}(9)_{\text{BL}}(10)_{\omega,\phi}[73]$ \\
    $\eta' \rightarrow \mu^+\mu^-$ & $3.11(5)_{\text{disp}}(7)_{\text{BL}}(9)_{\text{asym}}(1)_{\omega,\phi}[13]$& $ -5.626 (21)_{\text{disp}}(3)_{\text{BL}}(3)_{\omega,\phi}[22]$ \\
    \bottomrule
  \end{tabular}
  \caption{Final results for the real and imaginary part of the reduced amplitude for $\etapp$ decays into lepton pairs, including the $Z$-boson contribution~\eqref{SM_Z_num}. The uncertainties refer to the systematics of the dispersive representation (``disp''), the uncertainties propagated from the singly-virtual space-like data $Q^2\geq 5\GeV^2$ (``BL''), the asymptotic contribution (``asym''), and the weights in the isoscalar contribution (``$\omega,\phi$''). Total errors (square brackets) are added in quadrature.}\label{tab:ReImA}
  \renewcommand{\arraystretch}{1.0}
\end{table}

Putting together all contributions according to the TFF decomposition~\eqref{Eq:tff_compl}, we obtain the results for $\Re \A_\ell$ and $\Im \A_\ell$ for the four channels as summarized in Table~\ref{tab:ReImA}, where the uncertainties are propagated from the various sources and added in quadrature in the end. We emphasize that the dominant contributions to the imaginary part due to two-photon intermediate states are taken directly from Eq.~\eqref{Imgg}, as these are known with negligible uncertainty thanks to the normalization with respect to $F_{P\gamma\gamma}$. Sizable corrections to the two-photon contribution are observed for the $\eta^\prime$,
\begin{align}
\label{Im_parts}
\Imgg \mathcal{A}_e[\eta]&=-21.92 \to -21.84(0), &  
\Imgg \mathcal{A}_\mu[\eta]&=-5.47 \to -5.45(0),\notag\\
\Imgg \mathcal{A}_e[\eta^\prime]&=-23.68 \to -18.87(7), &  
\Imgg \mathcal{A}_\mu[\eta^\prime]&=-7.06 \to -5.63(2),
\end{align}
while for the $\eta$ the corrections come out about a factor of two larger than for $K_L$ decays~\cite{Hoferichter:2023wiy}
\beq
\Imgg \mathcal{A}_e[K_L]=-21.62 \to -21.59(1),\qquad    
\Imgg \mathcal{A}_\mu[K_L]=-5.21 \to -5.20(0).
\eeq

\begin{table}[t]
\renewcommand{\arraystretch}{1.3}
  \centering
 \scalebox{0.975}{  \begin{tabular}{lrr}
    \toprule
    Decay mode & $\Br[P\rightarrow \ell^+\ell^-]/\Br[P\rightarrow \gamma\gamma]$  & $\Br[P\rightarrow \ell^+\ell^-]$  \\
    \midrule 
    $\eta\rightarrow e^+e^-$ & $1.364(7)_{\text{disp}}(3)_{\text{BL}}(4)_{\text{asym}}(1)_{\omega,\phi}[9] \times 10^{-8}$ &  $5.37(4)_\text{red}(2)_\text{norm}[4]  \times 10^{-9}$  \\
    $\eta\rightarrow \mu^+\mu^-$ & $1.154(6)_{\text{disp}}(3)_{\text{BL}}(6)_{\text{asym}}(1)_{\omega,\phi}[9] \times 10^{-5}$ &  $4.54(4)_\text{red}(2)_\text{norm}[4]  \times 10^{-6}$  \\
    $\eta' \rightarrow e^+e^-$ & $7.809(28)_{\text{disp}}(52)_{\text{BL}}(29)_{\text{asym}}(6)_{\omega,\phi}[66] \times 10^{-9}$ & $1.80(2)_\text{red}(3)_\text{norm}[3]  \times 10^{-10}$ \\
    $\eta' \rightarrow \mu^+\mu^-$ & $5.294(10)_{\text{disp}}(57)_{\text{BL}}(74)_{\text{asym}}(4)_{\omega,\phi}[94] \times 10^{-6}$ &  $1.22(2)_\text{red}(2)_\text{norm}[3] \times 10^{-7}$  \\
    \bottomrule
  \end{tabular}}
  \caption{Final results for the normalized and total branching ratios for $\etapp$ decays into lepton pairs, with uncertainties labeled as in Table~\ref{tab:ReImA}. $\Br[\eta\rightarrow \gamma\gamma]=39.36(18)\%$, $\Br[\eta^\prime\rightarrow \gamma\gamma]=2.307(35)\%$ are taken from the RPP, and are known much more precisely than the normalizations from Eq.~\eqref{F_etagg_etapgg}, related to the two-photon widths via Eq.~\eqref{ggwidth}, because the uncertainty from the total widths drops out. For that reason, the resulting normalization errors (``norm'') for $\Br[P\rightarrow \ell^+\ell^-]$ are fairly small, of similar size as the ones derived from the reduced amplitude (``red'', corresponding to the total error in the first column).}
  \label{tab:Br}
  \renewcommand{\arraystretch}{1.0}
\end{table}

For the uncertainty estimate of the real part (and the non-$\gamma\gamma$ contributions to the imaginary part) we proceed as follows: firstly, for the uncertainty of the dispersive representation (``disp''), we observe that the integral cutoffs $\Lambda^2$ in Eq.~\eqref{Eq:A_disp}, previously varied between $2.25\GeV^2$ and $6.25 \GeV^2$ in accordance with the construction of the TFFs, do not suffice to reproduce the two-photon imaginary parts collected from all terms proportional to $L(0,0)$ in Eq.~\eqref{Eq:kernel_func}. This can be remedied by continuing the spectral functions according to  $\tilde\rho_P(x,y)\to \tilde\rho_P(\Lambda^2,y) \Lambda^2/x$ if one argument exceeds $\Lambda^2$, and according to~\cite{Holz:2024diw} 
\beq
\tilde \rho_P(x,y)\to \tilde \rho_P(\Lambda^2,\Lambda^2) \frac{\Lambda^2(x+\Lambda^2)(y+\Lambda^2)}{2xy(x+y)}
\eeq
if both arguments do, with the integral cutoffs now varied between $4\GeV^2$ and $10 \GeV^2$. Note that the dispersion integral of the decay amplitude into four pions contained in the double-spectral density was still evaluated up to cutoffs varied  between $2.25\GeV^2$ and $6.25 \GeV^2$. Breaking the correspondences between the formerly mentioned cutoffs and the ones employed in Eq.~\eqref{Eq:A_disp} then doubles the number of individual dispersive variants probed with respect to the TFF analysis in Refs.~\cite{Holz:2024lom,Holz:2024diw}. However, such an extension implies that the parameters of the effective poles need to be refit to ensure the normalization, which then allows us to obtain spectral functions that both fulfill the normalization condition and reproduce the correct two-photon imaginary parts. As in the TFF analysis, the effective coupling $g_\text{eff}$ in effective-pole variant $(A)$ comes out with negative sign for the $\eta$ and positive sign for the $\eta'$ with magnitudes observed $\lesssim 10 \, \%$. In general, the effective-pole contributions therefore remain reasonably small. Moreover, in the fits to high-energy singly-virtual TFF data, the mass scales observed for the effective mass parameter $M_\text{eff}$ remain compatible with the mass scales expected from contributions of higher intermediate states. Further, the dispersive representation of the $\etapp$ decay amplitude into four pions $f_1$, appearing in Eq.~\eqref{Eq:etapipig}, is matched onto the expected asymptotic behavior; see Ref.~\cite{Holz:2024diw} for more details. This matching is facilitated by a cut parameter $s_c$, which is varied between $1 \GeV^2$ and $1.5\GeV^2$. In the case of the $\eta$ decays, we additionally account for $U(3)$-symmetry violation by means of varying the coupling strength of the left-hand-cut contribution. The maximal variation of all probed dispersive variants with respect to the mean value is then assigned as the resulting dispersive uncertainty.

Secondly, the description of the TFF's singly-virtual Brodsky--Lepage limit (``BL'') proceeds via the effective-pole terms, the parameters of which, $M_\text{eff}$ in variant $(A)$ of Eq.~\eqref{Eq:TFF_eff} or one of the couplings $g_V$ in variant $(B)$, are varied within their fit uncertainty. In the variation of effective pole variant $(B)$ it is noteworthy that the overall TFF normalization is held fixed by readjusting the other respective coupling. Moreover, the spread of the respective observable when utilizing variants $(A)$ or $(B)$ is additionally accounted for in the BL uncertainty.

Thirdly, the uncertainty associated with the asymptotic contribution (``asym'') is evaluated by variation of the threshold parameter $s_\text{m}$, with $\sm=1.5(3)\GeV^2$ for the $\eta'$ and $\sm=1.4(4)\GeV^2$ for the $\eta$ as outlined in Sec.~\ref{sec:asym}. Furthermore, we consider the variation of the asymptotic coefficients $\bar{F}_\text{asym}^{P}$ in Eq.~\eqref{barF_asym_num} obtained via a superconvergence sum rule of the low-energy TFF representation with respect to the values of the lattice-QCD determination of Ref.~\cite{Bali:2021qem}. The resulting total asymptotic uncertainty only affects the real part of the reduced amplitudes. As shown in Table~\ref{tab:asymptotic}, the asymptotic contributions to $\Re\A_\ell$ only depend very weakly on $\ml$, so that, given the hierarchy between $\Re \A_e$ and $\Re\A_\mu$, the asymptotic uncertainties yield a much more important relative effect for the muon than for the electron channel.

Lastly, we consider the variation of the isoscalar weights of Eq.~\eqref{Eq:isosc_weights_values} within their respective error ranges for the uncertainty associated with the isoscalar $\omega,\phi$-contribution. While considering these variations, we readjust the parameters of the effective poles to hold the total TFF normalization fixed and ensure a realistic description of the high-energy singly-virtual TFF data. For both real and imaginary part of the reduced amplitudes this source of uncertainty turns out to be subleading, but is still listed for completeness.

Results for the branching fractions are provided in Table~\ref{tab:Br}, following the same procedure, which ensures that correlations between the uncertainties in the real and imaginary part are automatically taken into account. In the normalized case, the uncertainty of $F_{P\gamma\gamma}$ largely drops out, while the final prediction for $\Br[P\to\ell^+\ell^-]$ includes the normalization uncertainty via $\Br[\etapp\to\gamma\gamma]$. In this way, the sensitivity to $F_{P\gamma\gamma}$ is minimized, since the branching fractions $\Br[\etapp\to\gamma\gamma]$ do not require input for the total width of $\etapp$ and are therefore known much more precisely. 

\subsection{Comparison to previous work}

\begin{table}[t]
\renewcommand{\arraystretch}{1.3}
  \centering
  \begin{tabular}{lrrr}
    \toprule
    Decay mode & $\Re\mathcal{A}_\ell$ & $\Im\mathcal{A}_\ell$ &  $\Br[P\rightarrow \ell^+\ell^-]$ \\
    \midrule 
    $\eta\rightarrow e^+e^-$ & $(30.92$ ÷ $31.48) (11) $& $-21.92(0)$  & $(5.31$ ÷ $5.44 )(3)(2)(1)  \times 10^{-9}$ \\
    $\eta\rightarrow \mu^+\mu^-$ & $-(1.55$ ÷ $1.02)(5)$ & $-5.47(0)$ & $(4.72$ ÷ $4.52 )(2)(3)(4)  \times 10^{-6}$ \\
    $\eta' \rightarrow e^+e^-$ & $(47.43$ ÷ $48.23)(50)$ & $-21.0(5)$ & $(1.82$ ÷ $1.87 )(7)(2)(16)  \times 10^{-10}$ \\
    $\eta' \rightarrow \mu^+\mu^-$ & $(2.98$ ÷ $3.68)(19)$ & $-6.27(17)$ & $(1.36$ ÷ $1.49 )(5)(3)(25)  \times 10^{-7}$ \\
    \bottomrule
  \end{tabular}
  \caption{Results for the real and imaginary part of the reduced amplitude as well as the branching ratio as calculated in Ref.~\cite{Masjuan:2015cjl} using Canterbury approximants. The results span the range $a_{P;11}\in(2b_P^2, b_P^2)$ for an unknown doubly-virtual coefficient, corresponding to factorization and OPE constraints, where $b_P$ denotes the slope parameter of the TFF. For the reduced amplitude, the brackets are associated with the statistical error alone, while for the branching ratio, they represent the statistical error of $\etapp\rightarrow \gamma\gamma$, the uncertainty in $b_P$, and a systematic error. $\Br[\eta\to\gamma\gamma]=39.41(20)\%$, $\Br[\eta^\prime\to\gamma\gamma]=2.20(8)\%$  were taken from the 2014 RPP~\cite{ParticleDataGroup:2014cgo}.}
  \label{tab:masjuan_sanchez_puertas}
  \renewcommand{\arraystretch}{1.0}
\end{table}

\begin{table}[t]
\renewcommand{\arraystretch}{1.3}
  \centering
  \begin{tabular}{lrrr}
    \toprule
    Decay mode & This work & Ref.~\cite{Masjuan:2015cjl} & Experiment \\
    \midrule
    $\eta \rightarrow e^+e^-$ & $5.37(4) \times 10^{-9}$ & $(5.31$ ÷ $5.44 )(4)  \times 10^{-9}$  & $ < 7 \times 10^{-7}$  \cite{SND:2018egu} \\
    $\eta \rightarrow \mu^+\mu^-$ & $4.54(4) \times 10^{-6}$ & $(4.72$ ÷ $4.52 )(5)  \times 10^{-6}$ & $ 5.8(8) \times 10^{-6}$ \cite{ParticleDataGroup:2024cfk, Abegg:1994wx, Dzhelyadin:1980kj} \\
    $\eta^\prime \rightarrow e^+e^-$ & $ 1.80(3) \times 10^{-10}$ & $(1.82$ ÷ $1.87 )(18)  \times 10^{-10}$ & $ < 5.6 \times 10^{-9}$ \cite{Achasov:2015mek} \\
    $\eta^\prime \rightarrow \mu^+\mu^-$ & $1.22(3) \times 10^{-7}$ & $(1.36$ ÷ $1.49 )(26)  \times 10^{-7}$ & -- \\
    \bottomrule
  \end{tabular}
  \caption{Comparison of predicted branching ratios $\Br[P\to\ell^+\ell^-]$ from this work and from Ref.~\cite{Masjuan:2015cjl} with experiment. Note that for the $\eta^\prime$ channels using $\Br[\eta^\prime\to\gamma\gamma]$ from the 2024 RPP would increase the results from Ref.~\cite{Masjuan:2015cjl} by about $5\%$.}
  \label{tab:comparison}
  \renewcommand{\arraystretch}{1.0}
\end{table}

It is instructive to compare our results to previous work, e.g.,  the analysis from Ref.~\cite{Masjuan:2015cjl} using Canterbury approximants, which gives the most comprehensive study of $\etapp\to\ell^+\ell^-$ decays to date. In this case, the dominant systematic effect arises from the truncation in the Canterbury expansion, reflected by the ranges reproduced in  
Table~\ref{tab:masjuan_sanchez_puertas}. In addition, given that the rational approximation is designed for the space-like region, significant uncertainties affect the calculation of the imaginary part. For $\eta\to\ell^+\ell^-$, they are identified with the two-photon contributions~\eqref{Imgg}, while for $\eta^\prime\to\ell^+\ell^-$ an estimate of higher intermediate states is added. For the $\eta$ the additional contributions to the imaginary part are indeed small, but we obtain much larger corrections in the case of the $\eta^\prime$.\footnote{The results of the toy model from Ref.~\cite{Masjuan:2015cjl}, $\Im\A_e[\eta]=-21.805$, $\Im\A_\mu[\eta]=-5.441$, $\Im\A_e[\eta^\prime]=-19.251$, $\Im\A_\mu[\eta^\prime]=-5.733$, are much closer to our results given in Table~\ref{tab:ReImA}, but were not used in the final Canterbury evaluation.}

Similarly, the larger the pseudoscalar mass, the larger the corrections to a dispersion relation in the mass of the decaying particle~\cite{Dorokhov:2007bd,Dorokhov:2008cd} become, both because other intermediate states besides the two-photon cut increase in importance and because the TFF away from $q^2=M_P^2$ entails an uncontrolled dependence on the interpolating field. 
In the case of the $\pi^0$, evaluating the corresponding relation
\begin{align}
\label{Aq2}
 \Re\A_\ell(q^2)&=\A(0)+\frac{1}{\sigma_\ell(q^2)}\bigg[\text{Li}_2\big[-y_\ell(q^2)\big]
 +\frac{1}{4}\log^2\big[y_\ell(q^2)]+\frac{\pi^2}{12}\bigg],\\
 \A_\ell(0)&\simeq 3\log\frac{m_\ell}{\mu}-\frac{3}{2}\bigg[\int_0^{\mu^2}\diff t\frac{\tilde F_{P\gamma^*\gamma^*}(-t,-t)-1}{t}
 +\int_{\mu^2}^\infty \diff t\frac{\tilde F_{P\gamma^*\gamma^*}(-t,-t)}{t}\bigg]-\frac{5}{4},\notag
\end{align}
 leads to the change $\Re\A_e[\pi^0]=10.11(10)\to 9.95$~\cite{Hoferichter:2021lct},  suppressed by the small pseudoscalar mass as well as the dominance of the two-photon cut. Accordingly, for $\etapp$ decays the changes are much bigger 
 \begin{align}
  \Re\A_e[\eta]&=31.25(15)\to 27.80,&  \Re\A_\mu[\eta]&=-1.21(10)\to -2.07,\notag\\
 \Re\A_e[\eta^\prime]&=46.76(24)\to 37.59,&  \Re\A_\mu[\eta^\prime]&=3.11(13)\to 1.73, 
 \end{align}
 which demonstrates that for the heavier pseudoscalars a dispersion relation such as Eq.~\eqref{Aq2} breaks down completely.

Since, in all cases, the additional imaginary parts highlighted in Eq.~\eqref{Im_parts} reduce the magnitude of $\Im \mathcal{A}_\ell$, the resulting branching fractions come out smaller as well, see Table~\ref{tab:comparison}, albeit compatible within uncertainties. Besides the imaginary parts, also $\Re \mathcal{A}_\ell$ tends to take smaller values for all channels, which is likely related to a similar trend in the respective HLbL contributions~\cite{Holz:2024lom,Holz:2024diw,Masjuan:2017tvw}, since derived from the same set of $\etapp$ TFFs.\footnote{However, apart from $\eta^\prime\to e^+e^-$, our results for $\Re\A_\ell$ actually fall within the ranges from Ref.~\cite{Masjuan:2015cjl}. The doubly-virtual behavior of the TFF, which is not directly constrained by data (apart from a few data points in $\eta^\prime \to\gamma^*\gamma^*$ with large uncertainties~\cite{BaBar:2018zpn}), could be tested in future lattice-QCD calculations~\cite{ExtendedTwistedMass:2022ofm,Gerardin:2023naa}.} For the comparison to experiment, the tension in $\eta\to\mu^+\mu^-$ increases slightly from about $1.5\sigma$ to $1.6\sigma$, essentially due to the reduced uncertainty in our SM prediction, while for the other channels limits remain about two orders of magnitude away from the SM prediction, if available at all. Since our results now determine the SM predictions at the level of a few percent, this implies that in the $\etapp\to e^+e^-$ channels experiment could improve by about three orders of magnitude before theory uncertainties would need to be reconsidered. Meanwhile, in the $\eta\to\mu^+\mu^-$ channel our results again suggest a SM branching fraction slightly smaller than the current RPP average.\footnote{At the current level of precision, radiative corrections have not been considered in the experimental analyses, but, at least for the $\mu^+\mu^-$ channel, are expected to be much less critical than for $\pi^0\to e^+e^-$.}  A new measurement to clarify the situation appears well motivated, as could be possible within the REDTOP~\cite{REDTOP:2022slw} or other proposals for future high-statistics $\eta$ factories~\cite{Achasov:2023gey,Chen:2024wad,An:2025lws}. Similarly, it would be important to validate the experimental normalization~\eqref{F_etagg_etapgg}, a deficit in which could cause the tension in $\Br[\eta\to\mu^+\mu^-]$. Work in this direction is ongoing in the context of the JLab Primakoff program~\cite{Gan:2014pna}, addressing the currently inconclusive situation regarding previous Primakoff measurements~\cite{Browman:1974sj,Rodrigues:2008zza}.

\begin{table}[t]
\renewcommand{\arraystretch}{1.3}
  \centering
  \begin{tabular}{lrrrr}
    \toprule
    $P$&$\pi^0$ & $K_L$ & $\eta$ & $\eta^\prime$\\\midrule
    $P\to e^+e^-$ & $2.69(10)$  & $8.0(1.0)$ & $6.23(15)$ & $13.57(24)$\\
    $P\to\mu^+\mu^-$ & -- & $4.96(38)$ & $3.64(10)$ & $5.78(13)$\\
    \bottomrule
  \end{tabular}
  \caption{Low-energy constant $\chi^\text{r}(\mu)$, $\mu=0.77\GeV$, for the different pseudoscalar dilepton decays.}
  \label{tab:LEC}
  \renewcommand{\arraystretch}{1.0}
\end{table}

Finally, we can compare the results for $\Re \A_\ell$ to other pseudoscalar dilepton decays, taking out the leading structure-independent part by means of the one-loop expression in ChPT 
\beq
\label{Al_ChPT}
 \Re\A_\ell(q^2)|_\text{ChPT}=\frac{\text{Li}_2[-y_\ell(q^2)]+\frac{1}{4}\log^2\big[y_\ell(q^2)]+\frac{\pi^2}{12}}{\sigma_\ell(q^2)}+3\log\frac{m_\ell}{\mu}-\frac{5}{2}+\chi^{\text{r}}(\mu). 
\eeq
In this approximation, all structure-dependent effects are subsumed into the chiral low-energy constant $\chi^\text{r}(\mu)$. The results are summarized in Table~\ref{tab:LEC}, where the $Z$-boson contribution has been subtracted from the values of $\Re \A_\ell$ given in Table~\ref{tab:ReImA}, since Eq.~\eqref{Al_ChPT} corresponds to the long-range $\gamma^*\gamma^*$ diagram. From these results, it is obvious that a single low-energy constant does not suffice to describe all channels, both due to the residual dependence on the lepton  and the pseudoscalar mass, which lead to a substantial amount of violation of lepton flavor universality and $U(3)$ symmetry in $\chi^\text{r}(\mu)$, respectively.

\section{Constraints on physics beyond the Standard Model}
\label{sec:BSM}

\subsection{Effective operators}

Generalizing the corresponding expression from Ref.~\cite{Hoferichter:2021lct} (see also Ref.~\cite{Masjuan:2015cjl} for a similar analysis), the relevant effective operators are
\beq
\Lagr_\text{BSM}^{(1)}=\sum_{a=0,8}\bigg[C_{A,\ell}^{a} \bar q\frac{\lambda^a}{2}\gamma^\mu\gamma_5 q\, \bar \ell\gamma_\mu\gamma_5 \ell
+C_{P,\ell}^a \bar q\frac{\lambda^a}{2}i\gamma_5 q\, \bar \ell i\gamma_5 \ell\bigg]+C_{g,\ell}\frac{\alpha_s}{4\pi}G_{\mu\nu}^a\tilde G^{\mu\nu}_a\bar\ell i\gamma_5\ell,
\eeq
where instead of the triplet as for $\pi^0\to e^+e^-$ decay, now octet and singlet flavor components are being probed,\footnote{We use the standard conventions $\lambda^8=\frac{1}{\sqrt{3}}\text{diag}(1,1,-2)$, $\lambda^0=\sqrt{\frac{2}{3}}\mathds{1}$, and $\lambda^3=\text{diag}(1,-1,0)$ for the flavor decomposition. These operators can be matched onto SMEFT conventions~\cite{Grzadkowski:2010es,Buchmuller:1985jz}, cf.\ Ref.~\cite{Hoferichter:2021lct}, or, in a first step, identified with the corresponding operators in the low-energy EFT (LEFT).} and, in addition, a gluonic operator involving the field strength tensor $G_{\mu\nu}^a$ and its dual can play a role. For the matrix elements of these axial-vector, pseudoscalar, and gluonic operators, we use the conventions from Ref.~\cite{Hoferichter:2022mna}
\begin{align}
 \langle 0|\bar q \gamma^\mu\gamma_5 q|P(k)\rangle &= i b_q f_P^q k^\mu,\qquad  
 \langle 0|m_q\bar q i\gamma_5 q|P(k)\rangle = \frac{b_qh_P^q}{2},\notag\\ 
 \langle 0|\frac{\alpha_s}{4\pi} G^a_{\mu\nu} \tilde{G}_a^{\mu\nu}|P(k)\rangle&=a_P,\qquad b_u=b_d=\frac{1}{\sqrt{2}},\qquad b_s=1,
\end{align}
which are related by the Ward identity
\beq
b_q f_P^q M_P^2=b_q h_P^q-a_P,
\eeq
so that the final amplitude can be expressed in terms of $f_P^q$ and $a_P$. We obtain 
\begin{align}
 \Re \A_\ell(q^2)|_\text{BSM}&=-\frac{1}{\alpha^2 F_{P\gamma\gamma}}\bigg[\frac{C_{A,\ell}^8}{2\sqrt{3}}\Big(\frac{f_P^u+f_P^d}{\sqrt{2}}-2f_P^s\Big)
 +\frac{C_{A,\ell}^0}{2\sqrt{3}}\Big(f_P^u+f_P^d+\sqrt{2}f_P^s\Big)\\
&+\frac{M_P^2}{8\sqrt{3}\,\ml}\bigg(C_{P,\ell}^8\Big(\frac{f_P^u}{m_u\sqrt{2}}+\frac{f_P^d}{m_d\sqrt{2}}-2\frac{f_P^s}{m_s}\Big)
 +C_{P,\ell}^0\Big(\frac{f_P^u}{m_u}+\frac{f_P^d}{m_d}+\sqrt{2}\frac{f_P^s}{m_s}\Big)\bigg)\notag\\
&+\frac{a_P}{2\ml}\bigg(C_{g,\ell}+\frac{C_{P,\ell}^8}{4\sqrt{3}}\Big(\frac{1}{m_u}+\frac{1}{m_d}-\frac{2}{m_s}\Big)
+\frac{C_{P,\ell}^0}{2\sqrt{6}}\Big(\frac{1}{m_u}+\frac{1}{m_d}+\frac{1}{m_s}\Big)\bigg)
 \bigg].\notag
\end{align}
This expression reduces to the pion case by changing $C_{A,\ell}^8\to \sqrt{3}\,C_{A,\ell}^3$, $C_{P,\ell}^8\to \sqrt{3}\,C_{P,\ell}^3$ (as well as the relative sign of $f_P^d$), dropping singlet and gluonic terms, and specifying $f^u_\pi =-f^d_\pi= \sqrt{2}\, F_\pi$, $f^s_\pi=a_\pi=0$, i.e.,  \beq
\Re \A_\ell(q^2)|_\text{BSM}^\pi=-\frac{F_\pi}{\alpha^2 F_{\pi\gamma\gamma}}\bigg[C_{A,\ell}^3+\frac{M_{\pi^0}^2}{8\ml}\bigg(\frac{1}{m_u}+\frac{1}{m_d}\bigg)C_{P,\ell}^3\bigg],
\eeq
which coincides with Ref.~\cite{Hoferichter:2022mna} upon setting $m_u=m_d\equiv \hat m$. In the case of $\etapp$, further simplifications arise by converting the matrix elements into octet and singlet components 
\begin{align}
 	f^u_\eta &=f^d_\eta= \sqrt{\frac{2}{3}}F_8\cos\theta_8 - \frac{2}{\sqrt{3}} F_0\sin\theta_0,  &
 	f^s_\eta &= -\frac{2}{\sqrt{3}} F_8\cos\theta_8 - \sqrt{\frac{2}{3}} F_0\sin\theta_0,\notag  \\
 	f^u_{\eta^\prime} &=f^d_{\eta^\prime}= \sqrt{\frac{2}{3}}  F_8\sin\theta_8 + \frac{2}{\sqrt{3}} F_0\cos\theta_0, &
	f^s_{\eta^\prime} &= -\frac{2}{\sqrt{3}} F_8\sin\theta_8 + \sqrt{\frac{2}{3}}  F_0\cos\theta_0,
\end{align}
and by modifying the Lagrangian to include the quark mass matrix $\M_q=\text{diag}(m_u,m_d,m_s)$ in the definition of the pseudoscalar operator
\beq
\tilde \Lagr_\text{BSM}^{(1)}=\sum_{a=0,8}\tilde C_{P,\ell}^a \bar q \M_q\frac{\lambda^a}{2}i\gamma_5 q\, \bar \ell i\gamma_5 \ell,
\eeq
which, again for $m_u=m_d=\hat m$, amounts to setting 
\beq
\label{tilde_basis}
\hat m\big(\tilde C_P^8+\sqrt{2}\tilde C_P^0\big)=C_P^8+\sqrt{2} C_P^0,\qquad 
m_s\big(\sqrt{2}\tilde C_P^8-\tilde C_P^0\big)=\sqrt{2} C_P^8-C_P^0.
\eeq
In these conventions, the BSM contributions to $\etapp$ become
\begin{align}
\label{ReAell_BSM}
 \Re \A_\ell(q^2)|_\text{BSM}^\eta&=-\frac{1}{\alpha^2 F_{\eta\gamma\gamma}}\bigg[C_{A,\ell}^8 F_8\cos\theta_8-C_{A,\ell}^0 F_0\sin\theta_0+\frac{a_\eta}{2\ml}\Big(C_{g,\ell}+\frac{\sqrt{6}}{4}\tilde C_{P,\ell}^0\Big)\notag\\
 &+\frac{M_\eta^2}{4\ml}\Big(\tilde C_{P,\ell}^8 F_8\cos\theta_8-\tilde C_{P,\ell}^0 F_0\sin\theta_0\Big)\bigg],\notag\\
  \Re \A_\ell(q^2)|_\text{BSM}^{\eta^\prime}&=-\frac{1}{\alpha^2 F_{\eta^\prime\gamma\gamma}}\bigg[C_{A,\ell}^8 F_8\sin\theta_8+C_{A,\ell}^0 F_0\cos\theta_0+\frac{a_{\eta^\prime}}{2\ml}\Big(C_{g,\ell}+\frac{\sqrt{6}}{4}\tilde C_{P,\ell}^0\Big)\notag\\
 &+\frac{M_{\eta^\prime}^2}{4\ml}\Big(\tilde C_{P,\ell}^8 F_8\sin\theta_8+\tilde C_{P,\ell}^0 F_0\cos\theta_0\Big)\bigg].
\end{align}
As expected, $\etapp$ decays are primarily sensitive to the octet (singlet) component of the Wilson coefficient. Moreover, the pseudoscalar and gluonic terms display the expected chiral enhancement with $1/\ml$, while the chiral enhancement with the quark masses can be read off from Eq.~\eqref{tilde_basis}
\beq
\tilde C_{P,\ell}^8\simeq \frac{C_{P,\ell}^8+\sqrt{2}\,C_{P,\ell}^0}{3\hat m},\qquad
\tilde C_{P,\ell}^0\simeq \frac{\sqrt{2}\,C_{P,\ell}^8+2C_{P,\ell}^0}{3\hat m},
\eeq
where we expanded to leading order in $\hat m/m_s$. Accordingly, unless the Wilson coefficients contain explicit factors of the quark masses in specific BSM scenarios, the chiral enhancement in the quark mass is always sensitive to the lightest scale $\hat m$. 

Translating the experimental result for $\Br[\eta\to\mu^+\mu^-]$~\cite{ParticleDataGroup:2024cfk, Abegg:1994wx, Dzhelyadin:1980kj} to $\Re\A_\ell$, we have 
\beq
\label{eta_BSM}
 \Re\A_\mu(q^2)|_\text{exp}^{\eta}=(-3.18)(^{+1.03}_{-0.77}),
\eeq
where the uncertainty derives from the experimental branching fraction for the dilepton decay, while the uncertainties in $\Br[\eta\to\gamma\gamma]$ and the imaginary part are negligible. Accordingly, the tension with our SM prediction seems to increase from $1.6\sigma$ at the level of the branching fraction to $1.9\sigma$ at the level of $\Re\A_\ell$. However, due to the large relative uncertainty in $\Re\A_\ell$, linear error propagation is insufficient in this case, and, moreover, the error distribution becomes asymmetric, so that the $1.6\sigma$ interval of $\Re\A_\mu(q^2)|_\text{exp}^{\eta}$ really does include our SM prediction. As a result, we obtain the limit 
\beq
\label{etamumu_BSM}
 \big|\Re \A_\mu(q^2)|_\text{BSM}^\eta\big|<3.2\qquad \text{at } 90\% \text{ C.L.}
\eeq
In the other cases in which experimental limits are available, the sensitivity does not yet reach the SM prediction, and accordingly the BSM constraints are rather poor
\beq
\label{etapee_BSM}
\big|\Re \A_e(q^2)|_\text{BSM}^\eta\big|< 435,\qquad 
\big|\Re \A_e(q^2)|_\text{BSM}^{\eta^\prime}\big|< 280.
\eeq
Setting $C_{A,\ell}\simeq 1/\Lambda_{A,\ell}^2$, $\hat m\tilde C_{P,\ell}\simeq 1/\Lambda_{P,\ell}^2$, Eq.~\eqref{etamumu_BSM} probes scales $\Lambda_{A,\mu}\simeq 50\GeV$, $\Lambda_{P,\mu}\simeq 700\GeV$, the latter reflecting the chiral enhancement by $M_\eta/(2\sqrt{m_\mu\hat m})\simeq 14$. The same factor for the electron channels, $M_{\eta^\prime}/(2\sqrt{m_e\hat m})\simeq 360$, implies that Eq.~\eqref{etapee_BSM} actually probes scales as high as $\Lambda_{P,e}\simeq 1.6\TeV$, even though for the axial-vector operators the limits only reach a few GeV. In all cases we have suppressed the flavor indices, but due to 
\beq
\sqrt{\frac{F_8\cos\theta_8}{F_0\sin\theta_0}}\simeq 2.5,\qquad \sqrt{\frac{F_0\cos\theta_0}{F_8\sin\theta_8}}\simeq 1.6,
\eeq
the sensitivities to octet and singlet operators, when expressed in terms of the BSM scale, do not differ dramatically. 

\subsection{Light new particles}

Light new axial-vector ($Z'$) and pseudoscalar ($a$) states are described by the Lagrangian 
\beq
\Lagr^{(2)}_\text{BSM}=\sum_{f=\ell,q}\bar f\Big(c_A^f \gamma^\mu\gamma_5 Z_\mu'+c_P^f i\gamma_5 a\Big)f,
\eeq
which act as mediators between lepton and quark bilinears, with a result that can be expressed in terms of (momentum-dependent) Wilson coefficients 
\begin{align}
\label{Zprime_matching}
C_{A,\ell}^8&=-\frac{(c_A^u+c_A^d-2c_A^s)c_A^\ell}{\sqrt{3}\,M_{Z'}^2},& 
C_{P,\ell}^8&=\frac{(c_P^u+c_P^d-2c_P^s)c_P^\ell}{\sqrt{3}\,(m_a^2-q^2)},\notag\\
C_{A,\ell}^0&=-\sqrt{\frac{2}{3}}\frac{(c_A^u+c_A^d+c_A^s)c_A^\ell}{M_{Z'}^2},& 
C_{P,\ell}^0&=\sqrt{\frac{2}{3}}\frac{(c_P^u+c_P^d+c_P^s)c_P^\ell}{m_a^2-q^2}. 
\end{align}
For the axial-vector operators, the pole in the momentum transfer between quark and lepton bilinears cancels, while the pseudoscalar coefficients do display a pole at the mass of the mediator. Setting $c_A^u=-c_A^d=-c_A^s=-c_A^\ell=g/(4\cos\theta_W)$ and $M_{Z'}=M_Z$,  the axial-vector terms in Eq.~\eqref{ReAell_BSM} reproduce the $Z$-boson contribution in the SM~\eqref{SM_Z}. The leptonic couplings $c_{A,P}^\ell$ can be probed directly via the anomalous magnetic moment $a_\ell$, in which they produce a negative contribution~\cite{Leveille:1977rc}
 \begin{align}
  a_\ell^A&=-\frac{(c_A^\ell)^2m_\ell^2}{4\pi^2M_{Z'}^2}\int_0^1\diff x\frac{2x^3m_\ell^2+x(1-x)(4-x)M_{Z'}^2}{m_\ell^2 x^2+M_{Z'}^2(1-x)},\notag\\
  a_\ell^P&=-\frac{(c_P^\ell)^2m_\ell^2}{8\pi^2}\int_0^1\diff x\frac{x^3}{m_\ell^2 x^2+m_a^2(1-x)}.
 \end{align}
 Combining constraints from $a_\ell$ and $P\to\ell^+\ell^-$ thus allows one, in principle, to disentangle all flavor combinations in the case of the electron, while for the muon the triplet combination cannot be resolved because the $\pi^0$ channel is kinematically forbidden.   

\section{Conclusions}
\label{sec:summary}

In this work, we calculated the SM predictions for the dilepton decays of $\etapp$ based on a dispersive representation of the $\etapp\to\gamma^*\gamma^*$ TFFs, propagating the uncertainties from the various experimental input quantities, the systematics of the dispersion relation, and the matching to asymptotic constraints. For the latter, we derived a representation that allowed us to include pseudoscalar mass corrections in the same way as in our previous work on $\etapp$ pole contributions to HLbL scattering, so that the present predictions for the dilepton amplitudes and branching fractions, see Tables~\ref{tab:ReImA} and~\ref{tab:Br} for our main results, provide an implementation that takes full advantage of the detailed studies of the $\etapp$ TFFs in the HLbL context, leading to a precision at the few-percent level for all four branching fractions $\Br[\etapp\to\ell^+\ell^-]$. As a key advance over previous work, the dispersive approach allows for an improved account of the imaginary parts beyond the two-photon cut, which play an important role for the $\eta^\prime$ decays.

Confronting our SM prediction for $\eta\to\mu^+\mu^-$ with experiment, we confirmed a mild tension of $1.6\sigma$, already observed in earlier work using Canterbury approximants, emphasizing the importance of a new measurement of the $\eta\to\mu^+\mu^-$ channel as well as independent validation of the normalization from $\eta\to\gamma\gamma$ to clarify the situation.
In addition, further input on the doubly-virtual TFFs, from lattice QCD or experiment, would be valuable to cross check and refine the dispersive TFF representation.
The gain in precision largely originates from the real parts, as propagated from the TFFs via the loop integral, but for the $\eta^\prime$  we also find that higher intermediate states lead to a larger reduction in the magnitude of the imaginary part of the amplitude, translating to lower overall branching fractions. We provided an overview of the BSM limits that can be extracted from the currently available constraints, focusing on the chirality structure and the complementarity among different channels to resolve quark flavor components. Pseudoscalar (and gluonic) operators display a sizable chiral enhancement, thus probing large scales in the $\etapp\to e^+e^-$ channel even with the current experimental sensitivity for the branching fractions about two orders of magnitude above the SM prediction. Thanks to the level of precision achieved here, these limits could be improved by three orders of magnitude before theory uncertainties would need to be reconsidered.  
  
\paragraph{Note added.} While this paper was being finalized, a new preprint from BESIII became available~\cite{BESIII:2025jft}, giving $\Br[\eta\to\mu^+\mu^-]=5.8(1.0)(0.2)\times 10^{-6}$ and $\Br[\eta\to e^+e^-]< 2.2\times 10^{-7}$. The RPP average would change to $\Br[\eta\to\mu^+\mu^-]=5.8(6)\times 10^{-6}$, increasing the significance of the tension to the SM prediction to $2.1\sigma$.

\acknowledgments
Financial support by the Swiss National Science Foundation (Project Nos.\ 200020\_200553 and TMCG-2\_213690), the  Albert Einstein Center for
Fundamental Physics, and the DFG through the fund provided to the Research Unit  ``Photon--photon interactions in the Standard Model and beyond'' (Projektnummer 458854507 -- FOR 5327) is gratefully acknowledged.

\bibliographystyle{apsrev4-1_mod_2}
\bibliography{amu}

\end{document}